\documentstyle[12pt,cite,axodraw]{article}
\begin{document}

\title{\bf One-loop renormalization of general noncommutative
      Yang-Mills field model coupled to scalar and spinor fields}

\author{\sc I.L. Buchbinder${}^{a}$,
V.A. Krykhtin${}^b$
\footnote{e-mail: \tt joseph@tspu.edu.ru, krykhtin@mail2000.ru}
\\
\it ${}^a$Department of Theoretical Physics,\\
\it Tomsk State Pedagogical University,\\
\it Tomsk 634041, Russia\\
\it ${}^b$Laboratory of Mathematical Physics and\\
\it Department of Theoretical and Experimental Physics, \\
\it Tomsk Polytechnic University,\\
\it Tomsk 634050, Russia }

\date{}

\begin{titlepage}

\maketitle
\thispagestyle{empty}

\vspace{0.5cm}

\begin{abstract}
We study the theory of noncommutative $U(N)$ Yang-Mills field
interacting with scalar and spinor fields
in the fundamental and the adjoint representations.
We include in the action both the terms describing interaction
between the gauge and the matter fields
and
the terms which describe interaction among the
matter fields only.
Some of these interaction terms have not been considered
previously in the context of noncommutative field theory.
We find all counterterms for the theory to be finite in the one-loop
approximation.
It is shown that these counterterms
allow to
absorb all the divergencies by renormalization of the fields and the
coupling constants,
so the theory turns out to be multiplicatively renormalizable.
In case of 1PI gauge field functions
the result may easily be generalized on
an arbitrary number of the matter fields.
To generalize the results for the other 1PI functions it is
necessary for the
matter coupling constants to be adapted in the proper way.
In some simple cases this generalization for a part of these 1PI functions
is considered.
\end{abstract}

\end{titlepage}

\section{Introduction}

Noncommutative field theories have been attracting great attention for
the past few years.
Interest in these theories began with the discovery
of their relation to string theory (see \cite{SW} and references therein).
Apart from the string theory interest they are interesting on their own
as a sufficiently consistent non-local quantum field model.
(see reviews \cite{DN,ABGKM,Sz}).

Noncommutativity has some important consequences.
Two main consequencess are a restriction on the gauge group\footnote{We 
do not take into
account gauge groups which are only constructed perturbatively in the
noncommutativity parameter. Discussion of such field theories see i.g. in
\cite{0006246,0104153,0006091,0103209,0206003}.}
\cite{M,no-go}
and charge quantization \cite{H,0107055}.
One of the consequences at quantum level is the so-called UV/IR mixing.
Although the limit $\theta^{\mu\nu}\to{}0$ (\ref{theta})
($\theta^{\mu\nu}$ are the noncommutativity parameters)
reduces a classical noncommutative
theory to its commutative counterpart, at the quantum level this
is not the case due to the UV/IR mixing \cite{MRS,RS,ABK,0002075,0201144}.
This phenomenon of mixing of UV and IR singularities appears in the so-
called
nonplanar diagrams: some of the UV singularities of a commutative theory
convert in IR singularities in its noncommutative counterpart.
So,
contributions of the nonplanar diagrams to the effective action are 
singular
in $\theta^{\mu\nu}p_\nu$ ($p$ is external momenta).
These divergencies are
interpreted as IR ones \cite{MRS} and UV singularities  of the 
noncommutative
theories are not the same as in their commutative counterparts.
As a
consequence, it may violate renormalizability of the noncommutative field
theories.
Although  there is a general statement that a
noncommutative field theory should be renormalizable if its commutative
counterpart is
renormalizable
(see e.g. \cite{ChR} and the reviews \cite{DN,ABGKM,Sz})
we need an explicit check to support this statement in each
new concrete model
(see the discussion of this point in review \cite{ABGKM}).
By now,
as far as the nonsupersymmetric field theories are concerned, it has been
checked by direct calculations two-loop renormalizability
of $\phi^4_4$ theory \cite{BK,Micu} and one
loop renormalizability both pure noncommutative $U(N)$ gauge theory
\cite{A,BS,MS} and noncommutative $U(N)$ gauge theory interacting with
the fermionic field in the fundamental representation \cite{H} and
the bosonic field
in the adjoint representation \cite{ABKR} separately.
We
are going to consider here renormalizability of
a general theory of a noncommutative $U(N)$ gauge field interacting with 
matter fields.
But in contrast to the previous works where Yang-Mills field interacts 
with
only a single kind of matter field we
consider a most general action and
include the scalar and the spinor
fields both in the fundamental and in the adjoint representations.
The action also contains terms which describe interaction
among the matter fields and some of them have not been considered
previously.

Also we point out the activity concerning
the supersymmetric field theories.
There exist two approaches: the fermionic
coordinates of a superspace may be endowed with noncomutativity
\cite{FL} or not \cite{0104190}.
The second approach is more usual one.
Different quantum properties of matter and gauge fields have been
investigated both for ${\cal{}N}=1$
(see e.g. \cite{0002119,0107022,0109222})
and extended supersymmetric theories
(see e.g. \cite{0102007,0102045,0109130,0110203}).

This paper is organized as follows. In next section we briefly
review
basic properties of  noncommutative field theories and also fix the 
notation
and the action to be studied.
The action contains the scalar and the spinor
fields both in the fundamental and the adjoint representations
and
terms describing interaction among the fields.
In section 3 we find all counterterms needed to cancel
the divergencies of the
theory in the one-loop approximation.
It is shown that these counterterms allow us to carry out
renormalization of the fields and the coupling constants of the theory.
Thus, the theory is multiplicatively renormalizable in the one-loop 
approximation.
We also discuss the generalization of the theory for an arbitrary number 
of the matter fields.
In the Appendix we write out the propagators and the vertices of the
theory.
The calculations are given using the dimensional regularization
and standard methods of quantum field theory.
We do not consider the details of the calculations and
present only the final results.

\section{The Model}

We start this section with a brief formulation of some basic properties of
noncommutative field
theories.
As it well known that a noncommutative field theory may be
constructed from
commutative field theory by replacing the usual product of the fields by 
the
star one
\begin{eqnarray}
\label{theta}
f\cdot g\to
(f\star g)(x)=
  \left.\exp(\frac{i}{2}\theta^{\mu\nu}\partial_\mu^u\partial_\nu^v)
                f(x+u)g(x+v)\right|_{u=v=0} \neq (g\star f)(x),
\end{eqnarray}
where the constants $\theta^{\mu\nu}$ are the noncommutativity parameters.

As was shown in \cite{M,no-go}, the only possible gauge group admitting
simple noncommutative extension
(all pointwise products are replaced by the star one)
for a noncommutative gauge
field theory is $U(N)$.
Matter fields may transform or in the fundamental
representation\footnote{In principle, the matter fields can also belong to
antifundamental representation \cite{H}. However we do not consider this 
case
here.}
\begin{eqnarray*}
\phi'_i(x)=U^j_i(x)\star{}\phi_j(x),\qquad\qquad i,j=1,\ldots,N,
\end{eqnarray*}
or in the adjoint representation
\begin{eqnarray*}
\Phi'^i_j(x)=U^k_j(x)\star{}\Phi^m_k(x)\star{}U^+{}^i_m(x),
  \qquad\qquad  U^k_j\star{}U^+{}_k^i=U^+{}^k_j\star{}U^i_k=\delta^i_j.
\end{eqnarray*}
The covariant derivatives are defined as follows
\begin{eqnarray*}
D_\mu\phi_i&=&\partial_\mu\phi_i-igA_\mu{}^j_i\star{}\phi_j,
\\
D_\mu\Phi_j^i&=&\partial_\mu\Phi_j^i-igA_\mu{}^k_j\star{}\Phi_k^i
                                    +ig\Phi^k_j\star{}A_\mu{}^i_k
\\
 &\equiv&\partial_\mu\Phi_j^i-ig[A_\mu,\Phi]^i_j
\end{eqnarray*}
for the fundamental and the adjoint representations respectively.
Under the gauge
transformation these covariant derivatives transform as
\begin{eqnarray*}
D'_\mu\phi'_i&=&U^j_i\star{}D_\mu\phi_j,
\\
D'_\mu\Phi'^i_j&=&U^k_j\star{}D_\mu\Phi^m_k\star{}U^+{}^i_m
\end{eqnarray*}
if the gauge field $A$ has the transformation law
\begin{eqnarray}
\label{trA}
A'_\mu{}^i_j=U^k_j\star{}A_\mu{}^m_k\star{}U^+{}^i_m
     -\frac{i}{g}\partial_\mu U^k_j\star{}U^+{}^i_k.
\end{eqnarray}
As a consequence, the field strength takes the form
\begin{eqnarray*}
F_{\mu\nu}{}^i_j=\partial_\mu A_\nu{}^i_j-\partial_\nu A_\mu{}^i_j
       -igA_\mu{}_j^k\star{}A_\nu{}_k^i+igA_\nu{}_j^k\star{}A_\mu{}_k^i
\end{eqnarray*}
and has the following transformation law
\begin{eqnarray*}
F'_{\mu\nu}=U\star{}F_{\mu\nu}\star{}U^+.
\end{eqnarray*}
Hereafter we shall often omit matrix indices. From the
transformation law (\ref{trA}) we see $A$ may be restricted to be
selfconjugated $(A_\mu{}^i_j)^*=A_\mu{}^j_i$.

Now we can write down the action of the theory which we are going to study
\begin{eqnarray}
\label{IA}
S_{cl}&=&\int d^dx\left[
  \mbox{tr}\left[-\frac{1}{4}F_{\mu\nu}F^{\mu\nu}
  +{\bar\Psi}\star{}i\gamma^{\mu} D_{\mu}\Psi
  - M_{1}{\bar\Psi}\Psi\right]
 \right.
\\&{}&\mbox{}\nonumber
  +{\bar\psi}\star{}i\gamma^{\mu} D_{\mu}\psi
  - m_{1}{\bar\psi}\psi
\\&{}&\mbox{}\nonumber
  +D_{\mu}\phi^+\star{} D^{\mu}\phi
  -m^2_{2}\phi^+\phi
  -\frac{\lambda_1}{4!}\phi^+\star\phi\star\phi^+\star\phi
\\&{}&\mbox{}\nonumber
  +\mbox{tr}\left[ D_{\mu}\star{}\Phi^+ D^{\mu}\Phi
  -M^2_{2}\Phi^+\Phi\right]
\\&{}&\mbox{}\nonumber
  -\frac{\lambda_{2a}}{4!}\,
  \mbox{tr}\left[\Phi^+\star\Phi\star\Phi^+\star\Phi\right]
  -\frac{\lambda_{2b}}{4!}\,
  \mbox{tr}\left[\Phi^+\star\Phi^+\star\Phi\star\Phi\right]
\\&{}&\mbox{}\nonumber
  -\frac{\lambda_{3}}{4!}\,
  \mbox{tr}\left[\Phi\star\Phi\star\Phi\star\Phi\right]
  -\frac{\lambda^*_{3}}{4!}\,
  \mbox{tr}\left[\Phi^+\star\Phi^+\star\Phi^+\star\Phi^+\right]
\\&{}&\mbox{}\nonumber
    \left.
   -f_{a}\,\phi^{+}\star{}\Phi\star{}\Phi^{+}\star{}\phi
   -f_{b}\,\phi^{+}\star{}\Phi^{+}\star{}\Phi\star{}\phi
   -h\,{\bar\psi}\star{}\Psi\star{}\phi
   -h^{*}\,\phi^{+}\star{}{\bar\Psi}\star{}\psi
\,\,\right].
\end{eqnarray}
Here $\Psi$ is a fermionic field in the adjoint representation,
     $\psi$ is a fermionic field in the fundamental representation,
     $\Phi$ is a bosonic field in the adjoint representation and
     $\phi$ is a bosonic field in the fundamental representation.
In comparison with the works \cite{H,ABKR,Riad,AS} we have included in 
the action (\ref{IA})
scalar and spinor fields both in the fundamental and the adjoint
representations.
We have also included in the
action (\ref{IA}) terms
which describe interaction among the matter fields allowed by
symmetry and reality conditions.
Since in the literature only one matter field has been studied to be
coupled to a gauge field, the terms with $f_a$, $f_b$ and $h$ have never 
been
considered.
We use the couplings $f_a$, $f_b$, and $\lambda_{2a}$, $\lambda_{2b}$
as independent in contrast to
the works \cite{ABGKM,ABK,ABKR}.
Of course, we could include in the action (\ref{IA}) some more 
interaction terms (for
example $\Phi^+\Phi^3+\mbox{c.c.}$)
but in order to preserve multiplicative renormalizability
it is necessary to
consider in the action (\ref{IA}) mass-like terms containing
$\Phi^2+\mbox{c.c.}$
which would complicate a consideration.

The infinitesimal symmetry transformations of the action have the form
\begin{eqnarray*}
U&=&\exp{igT(x)}=1+igT(x)+\frac{1}{2}igT(x)\star igT(x)+\ldots
\\
\delta\psi&=&igT\star\psi,
   \qquad\qquad\delta\psi^+=-ig{\bar\psi}\star T,\qquad T^+=T,
\\
\delta\Psi&=&ig[T,\Psi],
   \qquad\qquad\delta{\bar\Psi}=ig[T,{\bar\Psi}],
\\
\delta\phi&=&igT\star\phi,\qquad\qquad\delta\phi^+=-ig\phi^+\star T,
\\
\delta\Phi&=&ig[T,\Phi],
   \qquad\qquad\delta\Phi^+=ig[T,\Phi^+],
\\
\delta A_\mu&=&\partial_\mu T(x)-ig[A_\mu,T],
\\
\delta F_{\mu\nu}&=&ig[T,F_{\mu\nu}].
\end{eqnarray*}
For any field $f$ one has
\begin{eqnarray*}
(\delta_{T_1}\delta_{T_2}-\delta_{T_2}\delta_{T_1})f
&=&\delta_{T_3}f,
\\
T_3&=&ig[T_1,T_2].
\end{eqnarray*}
We quantize this theory using the Faddeev-Popov method,
by introducing the ghost field
$C$ and antighost field ${\bar C}$, adding the
ghost action and the gauge-fixing term (we use the Lorentz gauge) to the 
initial
action (\ref{IA}). Then the action we quantize reads
\begin{eqnarray}
\label{theS}
S&=&S_{cl}+S_{GF+FP},
\\ \nonumber
S_{GF+FP}&=&-\int d^dx\,
       \mbox{tr}\left(\frac{1}{2\alpha}(\partial^\mu A_\mu)^2
       +{\bar C}\star{}\partial^\mu D_\mu C\right).
\end{eqnarray}

The aim of our analysis below is to calculate all one-loop divergencies 
and
to check the multiplicative renormalizability of the theory in the one-
loop approximation.
%%%%%%%%%%%%%%%%%%%%%%%%%%%%%%%%%%%%%%%%%%%%%%%%%%%%%%%%%%%%%%%%%
%%%%%%%%%%%%%%%%%%%%%%%%%%%%%%%%%%%%%%%%%%%%%%%%%%%%%%%%%%%%%%%%%
%%%%%%%%%%%%%%%%%%%%%%%%%%%%%%%%%%%%%%%%%%%%%%%%%%%%%%%%%%%%%%%%%
%%%%%%%%%%%%%%%%%%%%%%%%%%%%%%%%%%%%%%%%%%%%%%%%%%%%%%%%%%%%%%%%%
%%%%%%%%%%%%%%%%%%%%%%%%%%%%%%%%%%%%%%%%%%%%%%%%%%%%%%%%%%%%%%%%%
%%%%%%%%%%%%%%%%%%%%%%%%%%%%%%%%%%%%%%%%%%%%%%%%%%%%%%%%%%%%%%%%%

\section{Renormalization of the one-loop effective action}

Let $\Phi^A$ denote all the fields in the theory
$\Phi^A=(\phi,\phi^+,\Phi,\Phi^+,A_\mu,C,{\bar C},\psi,\bar
{\psi},\Psi,\bar{\Psi})$,
let  bosonic part of these
fields be $\varphi^i=(\phi,\phi^+,\Phi,\Phi^+,A_\mu)$
and fermionic part be
$\theta^\alpha=(C,{\bar C},\psi,\bar{\psi},\Psi,\bar{\Psi})$.
($A$, $i$ and $\alpha$ are condensed indices
which include discrete indices and space-time coordinates.
Both summing and integration over repeated indices are assumed.)
We use the background field method and split the action (\ref{theS})
into two parts $S_0$ and $V$, where $S_0$ is quadratic in its
fields and $V$ is the
rest of the total action (\ref{theS})
both depend on arbitrary background fields $\bar\Phi$ and quantum fields
$\Phi$.
Then, we have (up to a constant)
\begin{eqnarray*}
e^{i\Gamma_1}&=&\int D\Phi^A
   e^{\frac{i}{2}S_{AB}(\bar\Phi)\Phi^B\Phi^A},
\end{eqnarray*}
where $\Gamma_1$ is the one-loop effective action (EA)
and all derivatives in fields are right
$S_{AB}(\bar\Phi)=\frac{\displaystyle\delta_r}
                       {\vphantom{\Big|}\displaystyle\delta\bar\Phi{}^B}
                  \frac{\displaystyle\delta_rS(\bar\Phi)}
                       {\vphantom{\Big|}\displaystyle\delta\bar\Phi{}^A}
$\,.
We rewrite
$S_{AB}\Phi^B\Phi^A$ as
\begin{eqnarray*}
\frac{1}{2}S_{AB}(\bar\Phi)\Phi^B\Phi^A
&=&
          \frac{1}{2}S_{(ij)}(\bar\Phi)\varphi^i\varphi^j
          +\frac{1}{2}S_{[\beta\alpha]}(\bar\Phi)\theta^\alpha\theta^\beta
          +S_{(i\alpha)}(\bar\Phi)\theta^\alpha\varphi^i
\\
&=&
          \frac{1}{2}S_{(ij)}(\bar\Phi){\tilde\varphi}^i{\tilde\varphi}^j
                    +\frac{1}{2}{\tilde S}_{[\beta\alpha]}(\bar\Phi)
                                             \theta^\alpha\theta^\beta\, ,
\end{eqnarray*}
where
\begin{eqnarray*}
{\tilde\varphi^i}&=&\varphi^i+G^{ik}S_{k\alpha}\theta^\alpha,
                    \qquad S_{ij}G^{jk}=\delta^k_i     ,
\\
{\tilde S}_{[\beta\alpha]}&=&S_{[\beta\alpha]}
                 +G^{ij}S_{i[\alpha}S_{\beta]j}\, .
\end{eqnarray*}
Here $G^{ij}$, $S_{i\alpha}$ and $S_{\alpha\beta}$ depend on background
fields $\bar\Phi$.
After these redefinitions we get Gaussian functional integral and can 
integrate
over bosonic and fermionic fields respectively. As a result we have (up 
to a constant)
\begin{eqnarray}
\Gamma_1&=&\frac{i}{2}\mbox{\rm Tr}(\ln S_{ij}(\bar\Phi)-\ln S_{0ij})
              -\frac{i}{2}\mbox{\rm Tr}(\ln{\tilde S}_{[\beta\alpha]}
(\bar\Phi)
                -\ln{\tilde S}_{0[\beta\alpha]})\, .
\label{1ea}
\end{eqnarray}

Let us consider the first term in the rhs (\ref{1ea})
and do the following transformations
\begin{eqnarray}
\nonumber
\frac{i}{2}\,\mbox{\rm{}Tr}\,\left[\,\ln S_{ij}(\bar\Phi)-\ln S_{0ij}
\,\right]
  &=&
  \frac{i}{2}\,\mbox{\rm{}Tr}\,\left[\,\ln S_{0in}(\delta^n_j+G^{nk}_0V_
{kj}(\bar\Phi))
  -\ln\ S_{0ij}\,\right]
\\&=&\nonumber
  \frac{i}{2}\,\mbox{\rm{}Tr}\,\ln (\delta^i_j+G^{ik}_0V_{kj}(\bar\Phi))
\\&=&\label{sum}
  -\frac{i}{2}\,\sum^\infty_{n=1}
     \frac{(-1)^n}{n}\,\mbox{\rm{}Tr}\left(G^{ik}_0V_{kj}(\bar\Phi)\right)
^n\,.
\end{eqnarray}
Here $G_{0}^{ij}$ are propagators for the bosonic fields
$S_{0ij}G_0^{jk}=\delta^k_i$.
Then one can show by dimensional analisis, the divergences in
(\ref{sum}) may be originated only in the first four terms
\begin{eqnarray*}
\frac{i}{2}G^{ik}_0V_{ki},
&\qquad&
-\frac{i}{4}G^{ik}_0V_{kj}G^{jn}_0V_{ni},
\\
\frac{i}{6}G^{ik}_0V_{kj}G^{jn}_0V_{nm}G^{ml}_0V_{li},
&\qquad&
-\frac{i}{8}G^{ik}_0V_{kj}G^{jn}_0V_{nm}G^{ml}_0V_{lp}V_{pr}G^{rs}_0V_
{si},
\end{eqnarray*}
where $V_{ij}$ depends on backround fields $\bar\Phi$.
Doing similar procedure for the second term in the rhs of (\ref{1ea}) we 
find
that divergences may be originated in the following terms
\begin{eqnarray*}
-\frac{i}{2}G_0^{\alpha\beta}W_{\beta\alpha},
&\quad &
\frac{i}{4}G_0^{\alpha\beta}W_{\beta\gamma}
           G_0^{\gamma\delta}W_{\delta\alpha},
\\
-\frac{i}{6}G_0^{\alpha\beta}W_{\beta\gamma}
            G_0^{\gamma\delta}W_{\delta\varepsilon}
            G_0^{\varepsilon\sigma}W_{\sigma\alpha},
&\quad &
 \frac{i}{8}G_0^{\alpha\beta}W_{\beta\gamma}
            G_0^{\gamma\delta}W_{\delta\varepsilon}
            G_0^{\varepsilon\sigma}W_{\sigma\tau}
            G_0^{\tau\rho}W_{\rho\alpha}.
\end{eqnarray*}
Here $G_0^{\alpha\beta}$ are propagators for the fermionic fields
$S_{0\alpha\beta}G_0^{\beta\gamma}=\delta_\alpha^\gamma$ and
$W_{\beta\alpha}=V_{[\beta\alpha]}+G^{ij}V_{i[\alpha}V_{\beta]j}$
depending on background fields $\bar\Phi$.
To simplify calculations we perform the Fourier transformation of the
propagators and the vertices
\begin{eqnarray*}
G(x,x')&=&\int\left(\frac{dp}{2\pi}\right)^d
              \left(\frac{dp'}{2\pi}\right)^d
              e^{ipx+ip'x'}G(p,p')
         \equiv\int_{pp'}e^{ipx+ip'x'}G(p,p'),
\\[0.2cm]
  &&G(p,p')=\tilde{\delta}(p+p')G(p),
\\[0.2cm]
V(x,x')&=&\int_{pp'}e^{-ipx-ip'x'}V(p,p').
\end{eqnarray*}
Resulting propagators $G_0^{AB'}$ and verticies $V_{AB'}$
have been written out in Appendix \ref{FR}.

Let us review some more properties of the noncommutative field
theories. As it may easily be seen from the definition of the
star product (\ref{theta}), there is the following identity
\begin{eqnarray}
\label{free}
\int{}d^dx\, (\phi_1\star{}\phi_2)(x)=
 \int d^dx\, \phi_1(x)\phi_2(x)\,,
\end{eqnarray}
which is proved with the help of integration by parts and the
assumption that the functions $\phi_1(x)$ and $\phi_2(x)$ have
the proper asymptotic conditions. From this identity follows
that the quadratic part of the action for any noncommutative theory
is the same as that in its commutative counterpart. And as a
consequence the propagators of the noncommutative theory and the
commutative one coincide. The only thing which is modified is
the interaction. After Fourier transform of the fields
\begin{eqnarray*}
\phi(x)=
 \int\left(\frac{dp}{2\pi}\right)^d\,e^{ipx}\,\tilde{\phi}(p)
 \equiv\int_p e^{ipx}\,\tilde{\phi}(p)
\end{eqnarray*}
any interaction term gets an additional momentum dependence $V$
\begin{eqnarray}
\nonumber
&&\int d^dx (\phi_1\star{}\phi_2\star{}\ldots\star{}\phi_n)(x)
=
\int_{p_1\ldots{}p_n}\tilde{\delta}(p_1+p_2+\ldots+p_n)
  \tilde{\phi}(p_1)\ldots\tilde{\phi}(p_n)\,V(p_1,\ldots,p_n)
\\
\label{V}
&&V(p_1,\ldots,p_n)=
  e^{-\frac{i}{2}\sum_{j>i=1}^np_i\theta{}p_j}\,,
  \qquad
 p\theta{}k\equiv{}p_\mu\theta^{\mu\nu}k_\nu\,.
\end{eqnarray}
Due to this factor some diagrams become finite. Consider a
simple example of a one-loop graph. Let it contains two vertices
with three fields in each one
\begin{eqnarray*}
\int d^dx\, \phi_1{}^{i_1}_{i_3}\star{}
            \phi_2{}^{i_2}_{i_1}\star{}
            \phi_3{}^{i_3}_{i_2}(x)
\qquad
\mbox{and}
\qquad
\int d^dx'\, \phi_4{}^{j_1}_{j_3}\star{}
             \phi_5{}^{j_2}_{j_1}\star{}
             \phi_6{}^{j_3}_{j_2}(x')\,.
\end{eqnarray*}
Here $\phi$ are some fields and $i$ and $j$ are the group
indices. To get a one-loop graph from these vertices we need to
contract two fields from one vertex with two fields from another
one. With the help of the cyclic property of the star product
\begin{eqnarray*}
\int d^dx\, \phi_1\star{}\phi_2\star{}\ldots\star{}\phi_n
=
\int d^dx\, \phi_2\star{}\ldots\star{}\phi_n\star{}\phi_1
\end{eqnarray*}
which follows from (\ref{free}),
the first contraction may always be done between the last field
of the first vertex and the first field of the second one. In
momentum space this reads
\begin{eqnarray*}
&&\int_{p_1p_2p_3k_1k_2k_3}
  \tilde{\delta}(p_1+p_2+p_3)
  \tilde{\delta}(k_1+k_2+k_3)\,
  \tilde{\phi}_1{}^{i_1}_{i_3}(p_1)\,
  \tilde{\phi}_2{}^{i_2}_{i_1}(p_2)\times
\\&&\times
 <\tilde{\phi}_3{}^{i_3}_{i_2}(p_3)\,
  \tilde{\phi}_4{}^{j_1}_{j_3}(k_1)>
  \tilde{\phi}_5{}^{j_2}_{j_1}(k_2)\,
  \tilde{\phi}_6{}^{j_3}_{j_2}(k_3)\,
  V(p_1,p_2,p_3)\,
  V(k_1,k_2,k_3)\,.
\end{eqnarray*}
Any contraction of the fields has the following form
\begin{eqnarray*}
 <\tilde{\phi}_3{}^{i_3}_{i_2}(p_3)\,
  \tilde{\phi}_4{}^{j_1}_{j_3}(k_1)>
=\delta^{i_3}_{j_3}\,\delta^{j_1}_{i_2}
 \tilde{\delta}(p_3+k_1)\, G(p_3)\,.
\end{eqnarray*}
Integrating over $k_1$ and replacing $p_3\to{}p$ one gets
\begin{eqnarray}
\nonumber
&&\int_{pp_1p_2k_2k_3}
  \tilde{\delta}(p_1+p_2+p)
  \tilde{\delta}(-p+k_2+k_3)\,
  \tilde{\phi}_1{}^{i_1}_{i_3}(p_1)\,
  \tilde{\phi}_2{}^{i_2}_{i_1}(p_2)\,
  \tilde{\phi}_5{}^{j_2}_{i_2}(k_2)\,
  \tilde{\phi}_6{}^{i_3}_{j_2}(k_3)\times
\\&&
\nonumber
 \qquad\times\, V(p_1,p_2,p)\,
  V(-p,k_2,k_3)\,
  G(p)=
\\&&=
\nonumber
  \int_{p_1p_2k_2k_3}\tilde{\delta}(p_1+p_2+k_2+k_3)\,
  \tilde{\phi}_1{}^{i_1}_{i_3}(p_1)\,
  \tilde{\phi}_2{}^{i_2}_{i_1}(p_2)\,
  \tilde{\phi}_5{}^{j_2}_{i_2}(k_2)\,
  \tilde{\phi}_6{}^{i_3}_{j_2}(k_3)\,
  V(p_1,p_2,k_2,k_3)\times
\\&&\qquad\times
  \int_p\tilde{\delta}(p_1+p_2+p)\,G(p)
\label{1cont}
\end{eqnarray}
Here we use the equality
$V(p_1,\ldots,p_{n-1},p)V(-p,k_2,\ldots,k_n)=
 V(p_1,\ldots,p_{n-1},k_2,\ldots,k_n)$ which follows from the
 definition of $V$ (\ref{V}) and the delta functions
 $\tilde{\delta}(p_1+\ldots+p_{n-1}+p)$,
 $\tilde{\delta}(-p+k_2+\ldots+k_n)$.
Note that the group indices in (\ref{1cont}) are contracted as a
trace of product of all the fields.

To get a one-loop graph which is sufficient for our purpose, we
need one more contraction of the fields. This may be done by
several ways. If we contract neighbour fields (i.e. 2 and 5 or
1 and 6 assuming the cyclic property of the trace over the group
indices) we will get so-called "planar" diagram
which has the UV
divergences and its trace is over the product of all the fields
of the graph. Consider for example contraction of fields 1 and
6.
One has
\begin{eqnarray*}
 <\tilde{\phi}_1{}^{i_1}_{i_3}(p_1)\,
  \tilde{\phi}_6{}^{i_3}_{j_2}(k_3)>
=\delta^{i_1}_{j_2}\,\delta^{i_3}_{i_3}
 \tilde{\delta}(p_1+k_3)\, G'(p_1)\,.
\end{eqnarray*}
and after integration over $k_3$ and replacement $p_1\to{}p'$
\begin{eqnarray}
\nonumber
&&
 \delta^{i_3}_{i_3}
 \int_{p'p_2k_2}
 \tilde{\delta}(p_2+k_2)\,
 \tilde{\phi}_2{}^{i_2}_{i_1}(p_2)\,
 \tilde{\phi}_5{}^{i_1}_{i_2}(k_2)\,
 V(p',p_2,k_2,-p')\,\times
\\&&\qquad\qquad\times
 \int_{p}
 \tilde{\delta}(p_2+p+p')\,
 G(p)
 G'(p')=
\nonumber
\\
\nonumber
&&=
 N\int_{p_2k_2}
 \tilde{\delta}(p_2+k_2)\,
 \tilde{\phi}_2{}^{i_2}_{i_1}(p_2)\,
 \tilde{\phi}_5{}^{i_1}_{i_2}(k_2)\,
 V(p_2,k_2,)\,\times
\\&&\qquad\qquad\times
 \int_{pp'}
 \tilde{\delta}(p_2+p+p')\,
 G(p)
 G'(p')
\label{planar}
\end{eqnarray}
The last line of (\ref{planar}) is a usual one-loop UV divergent
integral.

Another variant of contraction of the fields are 1 and 5 or 2 and
6. In these cases we get so-called "nonplanar" diagrams. Let us
contract the fields 2 and 6.
One has
\begin{eqnarray*}
 <\tilde{\phi}_2{}^{i_2}_{i_1}(p_2)\,
  \tilde{\phi}_6{}^{i_3}_{j_2}(k_3)>
=\delta^{i_2}_{j_2}\,\delta^{i_3}_{i_1}
 \tilde{\delta}(p_2+k_3)\, G''(p_2)\,.
\end{eqnarray*}
After integration over $k_3$ and replacement $p_2\to{}p'$ we
have
\begin{eqnarray}
\nonumber
&&
 \int_{p'p_1k_2}
 \tilde{\delta}(p_1+k_2)\,
 \tilde{\phi}_1{}^{i_1}_{i_1}(p_1)\,
 \tilde{\phi}_5{}^{i_2}_{i_2}(k_2)\,
 V(p_1,p',k_2,-p')\,\times
\\&&\qquad\qquad\times
 \int_{p}
 \tilde{\delta}(p_1+p+p')\,
 G(p)
 G''(p')=
\nonumber
\\
\nonumber
&&=
 \int_{p_1k_2}
 \tilde{\delta}(p_1+k_2)\,
 \tilde{\phi}_1{}^{i_1}_{i_1}(p_1)\,
 \tilde{\phi}_5{}^{i_2}_{i_2}(k_2)\,
 V(p_1,k_2,)\,\times
\\&&\qquad\qquad\times
 \int_{pp'}
 \tilde{\delta}(p_1+p+p')\,
 G(p)
 G''(p')
 e^{ip'\theta{}k_2}
\label{nonplanar}
\end{eqnarray}
Here we see two features differing planar 
diagrams (\ref{planar}) from nonplanar ones (\ref{nonplanar}). The first
feature is the presence of the exponential factor
$e^{ip'\theta{}k_2}$ in (\ref{nonplanar}). Namely this factor
makes nonplanar diagrams finite. And the second feature is
that in the nonplanar diagrams the group indices trace is not
over the product of all the fields of the diagram. Due to this
we shall have for example equations like
(\ref{NPAA},\ref{NPAAA1},\ref{NPAAA2}).
More detailed information on this subject may be found i.g. in
refs. \cite{MRS,ChR,F}.

\subsection{Two-point gauge field function}
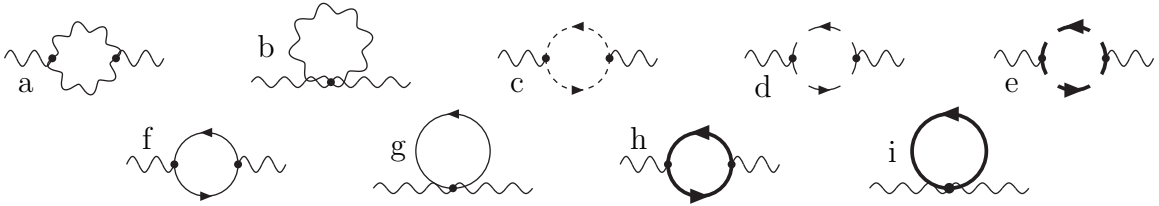
\begin{figure}[htbp]
\begin{picture}(455,65)(0,15)
\PhotonArc(41,57)(12,0,360){2}{8}
\Photon(11,57)(29,57){3}{2}
\Photon(53,57)(71,57){3}{2}
\Vertex(29,57){1.5}
\Vertex(53,57){1.5}
\Text(19,47)[]{\mbox{}a}
%---------------------------------------------------
\Photon(134,48)(104,48){2}{3}
\Photon(134,48)(164,48){2}{3}
\PhotonArc(134,62)(14,-90,270){2}{8}
\Vertex(134,48){1.5}
\Text(110,62)[]{\mbox{}b}
%---------------------------------------------------
\DashArrowArc(227,57)(12,0,180){2}
\DashArrowArc(227,57)(12,180,360){2}
\Photon(197,57)(215,57){3}{2}
\Photon(239,57)(257,57){3}{2}
\Vertex(215,57){1.5}
\Vertex(239,57){1.5}
\Text(205,47)[]{\mbox{}c}
%---------------------------------------------------
\DashArrowArc(320,57)(12,0,180){7}
\DashArrowArc(320,57)(12,180,360){7}
\Photon(290,57)(308,57){3}{2}
\Photon(332,57)(350,57){3}{2}
\Vertex(308,57){1.5}
\Vertex(332,57){1.5}
\Text(298,47)[]{\mbox{}d}
%---------------------------------------------------
\SetWidth{1.4}
\DashArrowArc(414,57)(12,0,180){7}
\DashArrowArc(414,57)(12,180,360){7}
\SetWidth{0.5}
\Photon(384,57)(402,57){3}{2}
\Photon(426,57)(444,57){3}{2}
\Vertex(402,57){1.5}
\Vertex(426,57){1.5}
\Text(392,47)[]{\mbox{}e}
%---------------------------------------------------
%---------------------------------------------------
\ArrowArc(87,17)(12,0,180)
\ArrowArc(87,17)(12,180,360)
\Photon(57,17)(75,17){3}{2}
\Photon(99,17)(117,17){3}{2}
\Vertex(75,17){1.5}
\Vertex(99,17){1.5}
\Text(65,27)[]{\mbox{}f}
%---------------------------------------------------
\Photon(180,8)(150,8){2}{3}
\Photon(180,8)(210,8){2}{3}
\ArrowArc(180,22)(14,-90,270)
\Vertex(180,8){1.5}
\Text(160,22)[]{\mbox{}g}
%---------------------------------------------------
%---------------------------------------------------
\SetWidth{1.4}
\ArrowArc(273,17)(12,0,180)
\ArrowArc(273,17)(12,180,360)
\SetWidth{0.5}
\Photon(243,17)(261,17){3}{2}
\Photon(285,17)(303,17){3}{2}
\Vertex(261,17){1.5}
\Vertex(285,17){1.5}
\Text(251,27)[]{\mbox{}h}
%---------------------------------------------------
\Photon(367,8)(337,8){2}{3}
\Photon(367,8)(397,8){2}{3}
\SetWidth{1.4}
\ArrowArc(367,22)(14,-90,270)
\SetWidth{0.5}
\Vertex(367,8){2.0}
\Text(347,22)[]{\mbox{}i}
\end{picture}
\caption{Diagrams contributing to the two-point gauge field function}
\label{2Ph}
\end{figure}

The diagrams which give the one-loop correction to the gauge field self-
energy are
shown in Figure~\ref{2Ph}.
Note that we may generalize the consideration of the two-point gauge field
function to an arbitrary number of the matter fields.
Let
 $n_f$ be the number of the fermionic fields in the fundamental
 representation\footnote{More precisely, $n_f$ is the number of 
multiplets (N
 fields in each) of the fermionic fields in the fundamental 
representation.
 Using QCD terminology, $n_f$ is the number of flavours, N is the number 
of
 colours. For the other fields situation is similar.},
 $n_F$ be the number of the fermionic fields in the adjoint 
representation,
 $n_b$ be the number of the bosonic fields in the fundamental 
representation,
 $n_B$ be the number of the bosonic fields in the adjoint representation.
The tadpole diagram with a gauge field loop
(Fig.\ref{2Ph}b) has no UV
divergence.
Using the minimal substraction scheme and the dimensional regularization 
we
find that the other
diagrams give the following contributions to the one-loop counterterm
\begin{eqnarray}
\nonumber
S_{1A^2}&=&\frac{1}{(4\pi)^2}\frac{-\frac{1}{6}g^2}{d-4}
   \int d^dx\, \mbox{tr}\left[
   (\partial_\mu A_\nu-\partial_\nu A_\mu)
			  (\partial^\mu A^\nu-\partial^\nu A^\mu)
              \right]\times
\\&&{}
  \label{AA}
       \begin{array}{lcccccr}
       {}&a+c&d&e&f+g&h+i&{}\\
       \times\left[\right.&
       N(3\alpha-13)&+4n_f&+8Nn_F&+n_b&+2Nn_B
       &\left.\right].
       \end{array}
\end{eqnarray}
As a consequence, the renormalizations of $A$ and $\alpha$ are easily 
found
\begin{eqnarray}
\stackrel{\circ}{A_\mu}&=&Z_AA_\mu
  \qquad Z_A=1+\frac{1}{(4\pi)^2}\frac{\frac{1}{3}g^2}{d-4}
         \left[n_b+4n_f+N(3\alpha-13+2n_B+8n_F)\right],
\label{A}
\\
\stackrel{\circ}{\alpha}&=&Z_\alpha \alpha\qquad
     Z_\alpha=Z_A^2.
\nonumber
\end{eqnarray}
Here the bare quantities are labeled with $\circ$ mark.
From (\ref{A}) we see that the renormalization of $SU(N)$ part
of the gauge fields is the same as in commutative $SU(N)$ gauge
theory with the same matter field content.

%The renormalization of $U(1)$ part of the gauge fields are not
%the same as in commutative QED due to
%the appearance of interactions of it
%both with $SU(N)$ and $U(1)$ parts.

Also note here that the non-planar contributions of these diargams have 
the
following structure
\begin{eqnarray}
\int_{k_1k_2}\tilde\delta(k_1+k_2)\
      \mbox{tr}\tilde A_{\alpha}(k_1)\
      \mbox{tr}\tilde A_{\beta}(k_2)
  \int_{k}f^{\alpha\beta}(k,k_1,k_2),
\label{NPAA}
\end{eqnarray}
where $f^{\alpha\beta}(k,k_1,k_2)$ are some functions. If we denote $T_0$ 
and
$T_a$ to be generators of $U(1)$ and $SU(N)$ groups respectively
$(U(N)=U(1)\times{}SU(N))$, then we will see that only $U(1)$ part (and 
not
$SU(N)$ part) of $U(N)$ group contributes to (\ref{NPAA}) due to
tracelessness of $T_a$.
So term (\ref{NPAA}) with UV/IR mixing depends on $U(1)$ part of
the gauge fields only.

%%%%%%%%%%%%%%%%%%%%%%%%%%%%%%%%%%%%%%%%%%%%%%%%%%%%%%%

\subsection{Three- and four-point gauge field functions}
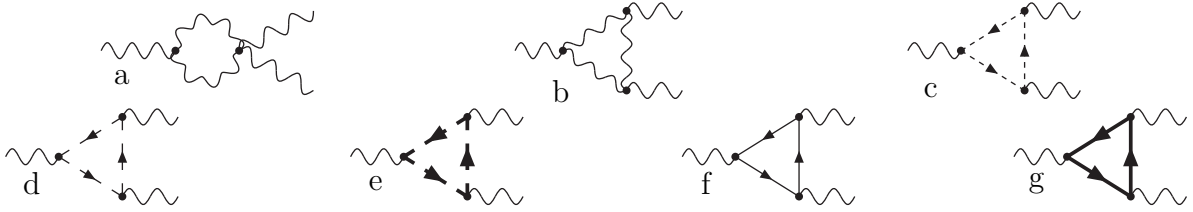
\begin{figure}[htbp]
\begin{picture}(455,65)(0,05)
\PhotonArc(81,57)(12,0,360){2}{8}
\Photon(41,57)(69,57){3}{3}
\Photon(93,57)(121,72){3}{3}
\Photon(93,57)(121,42){3}{3}
\Vertex(69,57){1.5}
\Vertex(93,57){1.5}
\Text(49,47)[]{\mbox{}a}
%---------------------------------------------------
\Photon(215,57)(239,42){2}{3}
\Photon(239,42)(239,72){2}{3}
\Photon(239,72)(215,57){2}{3}
\Photon(197,57)(215,57){3}{2}
\Photon(239,42)(260,42){3}{2}
\Photon(239,72)(260,72){3}{2}
\Vertex(215,57){1.5}
\Vertex(239,72){1.5}
\Vertex(239,42){1.5}
\Text(215,42)[]{\mbox{}b}
%---------------------------------------------------
\DashArrowLine(365,57)(389,42){2}
\DashArrowLine(389,42)(389,72){2}
\DashArrowLine(389,72)(365,57){2}
\Photon(345,57)(365,57){3}{2}
\Photon(389,42)(410,42){3}{2}
\Photon(389,72)(410,72){3}{2}
\Vertex(365,57){1.5}
\Vertex(389,72){1.5}
\Vertex(389,42){1.5}
\Text(355,42)[]{\mbox{}c}
%---------------------------------------------------
%---------------------------------------------------
\DashArrowLine(25,17)(49,2){7}
\DashArrowLine(49,2)(49,32){7}
\DashArrowLine(49,32)(25,17){7}
\Photon(5,17)(25,17){3}{2}
\Photon(49,2)(70,2){3}{2}
\Photon(49,32)(70,32){3}{2}
\Vertex(25,17){1.5}
\Vertex(49,32){1.5}
\Vertex(49,2){1.5}
\Text(15,7)[]{\mbox{}d}
%---------------------------------------------------
\SetWidth{1.4}
\DashArrowLine(155,17)(179,2){7}
\DashArrowLine(179,2)(179,32){7}
\DashArrowLine(179,32)(155,17){7}
\SetWidth{0.5}
\Photon(135,17)(155,17){3}{2}
\Photon(179,2)(200,2){3}{2}
\Photon(179,32)(200,32){3}{2}
\Vertex(155,17){1.5}
\Vertex(179,32){1.5}
\Vertex(179,2){1.5}
\Text(145,7)[]{\mbox{}e}
%---------------------------------------------------
\ArrowLine(280,17)(304,2)
\ArrowLine(304,2)(304,32)
\ArrowLine(304,32)(280,17)
\Photon(260,17)(280,17){3}{2}
\Photon(304,2)(325,2){3}{2}
\Photon(304,32)(325,32){3}{2}
\Vertex(280,17){1.5}
\Vertex(304,32){1.5}
\Vertex(304,2){1.5}
\Text(270,7)[]{\mbox{}f}
%---------------------------------------------------
\SetWidth{1.4}
\ArrowLine(405,17)(429,2)
\ArrowLine(429,2)(429,32)
\ArrowLine(429,32)(405,17)
\SetWidth{0.5}
\Photon(385,17)(405,17){3}{2}
\Photon(429,2)(450,2){3}{2}
\Photon(429,32)(450,32){3}{2}
\Vertex(405,17){1.5}
\Vertex(429,32){1.5}
\Vertex(429,2){1.5}
\Text(395,7)[]{\mbox{}g}
\end{picture}
\caption{Diagrams contributing to the three-point gauge field function}
\label{3Ph}
\end{figure}

\begin{figure}[htbp]
\begin{picture}(455,110)(0,0)
%-----------------------------------------
\Vertex(50,92){1.5}
\Vertex(74,92){1.5}
\PhotonArc(62,92)(12,0,360){2}{10}
\Photon(50,92)(36,104){2}{3}
\Photon(50,92)(36,80){2}{3}
\Photon(74,92)(88,104){2}{3}
\Photon(74,92)(88,80){2}{3}
\Text(36,92)[]{\mbox{}a}
%-----------------------------------------
\Vertex(178,104){1.5}
\Vertex(202,104){1.5}
\Vertex(190,80){1.5}
\Photon(202,104)(178,104){2}{3}
\Photon(178,104)(190,80){2}{3}
\Photon(202,104)(190,80){2}{3}
\Photon(164,80)(190,80){2}{4}
\Photon(216,80)(190,80){2}{4}
\Photon(164,104)(178,104){2}{2}
\Photon(202,104)(216,104){2}{2}
\Text(164,92)[]{\mbox{}b}
%-----------------------------------------
\Vertex(295,104){1.5}
\Vertex(319,104){1.5}
\Vertex(295,80){1.5}
\Vertex(319,80){1.5}
\Photon(295,104)(319,104){2}{3}
\Photon(319,104)(319,80){2}{3}
\Photon(319,80)(295,80){2}{3}
\Photon(295,80)(295,104){2}{3}
\Photon(280,104)(295,104){2}{2}
\Photon(319,104)(334,104){2}{2}
\Photon(280,80)(295,80){2}{2}
\Photon(319,80)(334,80){2}{2}
\Text(280,92)[]{\mbox{}c}
%-----------------------------------------
\Vertex(401,104){1.5}
\Vertex(425,104){1.5}
\Vertex(401,80){1.5}
\Vertex(425,80){1.5}
\DashArrowLine(401,104)(425,104){2}
\DashArrowLine(425,104)(425,80){2}
\DashArrowLine(425,80)(401,80){2}
\DashArrowLine(401,80)(401,104){2}
\Photon(401,104)(386,104){2}{2}
\Photon(425,104)(440,104){2}{2}
\Photon(401,80)(386,80){2}{2}
\Photon(425,80)(440,80){2}{2}
\Text(386,92)[]{\mbox{}d}
%-----------------------------------------
%-----------------------------------------
\Vertex(50,40){1.5}
\Vertex(50,64){1.5}
\Vertex(74,40){1.5}
\Vertex(74,64){1.5}
\DashArrowLine(50,40)(50,64){5}
\DashArrowLine(50,64)(74,64){5}
\DashArrowLine(74,64)(74,40){5}
\DashArrowLine(74,40)(50,40){5}
\Photon(50,40)(36,40){2}{2}
\Photon(50,64)(36,64){2}{2}
\Photon(74,40)(88,40){2}{2}
\Photon(74,64)(88,64){2}{2}
\Text(36,52)[]{\mbox{}e}
%-----------------------------------------
\Vertex(178,40){1.5}
\Vertex(178,64){1.5}
\Vertex(202,40){1.5}
\Vertex(202,64){1.5}
\SetWidth{1.4}
\DashArrowLine(178,40)(178,64){5}
\DashArrowLine(178,64)(202,64){5}
\DashArrowLine(202,64)(202,40){5}
\DashArrowLine(202,40)(178,40){5}
\SetWidth{0.5}
\Photon(178,40)(164,40){2}{2}
\Photon(202,40)(216,40){2}{2}
\Photon(178,64)(164,64){2}{2}
\Photon(202,64)(216,64){2}{2}
\Text(164,52)[]{\mbox{}f}
%-----------------------------------------
\Vertex(295,52){1.5}
\Vertex(319,52){1.5}
\ArrowArc(307,52)(12,0,180)
\ArrowArc(307,52)(12,180,360)
\Photon(295,52)(280,64){2}{3}
\Photon(295,52)(280,40){2}{3}
\Photon(319,52)(334,40){2}{3}
\Photon(319,52)(334,64){2}{3}
\Text(280,52)[]{\mbox{}g}
%-----------------------------------------
\Vertex(401,64){1.5}
\Vertex(425,64){1.5}
\Vertex(413,40){1.5}
\Photon(413,40)(440,40){2}{4}
\Photon(413,40)(386,40){2}{4}
\Photon(401,64)(386,64){2}{2}
\Photon(425,64)(440,64){2}{2}
\ArrowLine(401,64)(425,64)
\ArrowLine(425,64)(413,40)
\ArrowLine(413,40)(401,64)
\Text(386,52)[]{\mbox{}h}
%-----------------------------------------
%-----------------------------------------
\Vertex(74,0){1.5}
\Vertex(50,0){1.5}
\Vertex(50,24){1.5}
\Vertex(74,24){1.5}
\ArrowLine(50,0)(74,0)
\ArrowLine(74,0)(74,24)
\ArrowLine(74,24)(50,24)
\ArrowLine(50,24)(50,0)
\Photon(50,0)(36,0){2}{2}
\Photon(74,0)(88,0){2}{2}
\Photon(50,24)(36,24){2}{2}
\Photon(74,24)(88,24){2}{2}
\Text(36,12)[]{\mbox{}i}
%-----------------------------------------
\SetWidth{1.4}
\ArrowArc(190,12)(12,0,180)
\ArrowArc(190,12)(12,180,360)
\SetWidth{0.5}
\Vertex(178,12){1.5}
\Vertex(202,12){1.5}
\Photon(202,12)(220,0){2}{3}
\Photon(202,12)(220,24){2}{3}
\Photon(178,12)(160,24){2}{3}
\Photon(178,12)(160,0){2}{3}
\Text(164,12)[]{\mbox{}j}
%-----------------------------------------
\Vertex(307,0){1.5}
\Vertex(319,24){1.5}
\Vertex(295,24){1.5}
\Photon(307,0)(280,0){2}{4}
\Photon(307,0)(334,0){2}{4}
\Photon(280,24)(295,24){2}{2}
\Photon(334,24)(319,24){2}{2}
\SetWidth{1.4}
\ArrowLine(307,0)(319,24)
\ArrowLine(319,24)(295,24)
\ArrowLine(295,24)(307,0)
\SetWidth{0.5}
\Text(280,12)[]{\mbox{}k}
%-----------------------------------------
\Vertex(425,0){1.5}
\Vertex(425,24){1.5}
\Vertex(401,0){1.5}
\Vertex(401,24){1.5}
\Photon(425,24)(440,24){2}{2}
\Photon(425,0)(440,0){2}{2}
\Photon(386,0)(401,0){2}{2}
\Photon(386,24)(401,24){2}{2}
\SetWidth{1.4}
\ArrowLine(425,0)(425,24)
\ArrowLine(425,24)(401,24)
\ArrowLine(401,24)(401,0)
\ArrowLine(401,0)(425,0)
\SetWidth{0.5}
\Text(386,12)[]{\mbox{}l}
\end{picture}
\caption{Diagrams contributing to the four-point gauge field function}
\label{4Ph}
\end{figure}
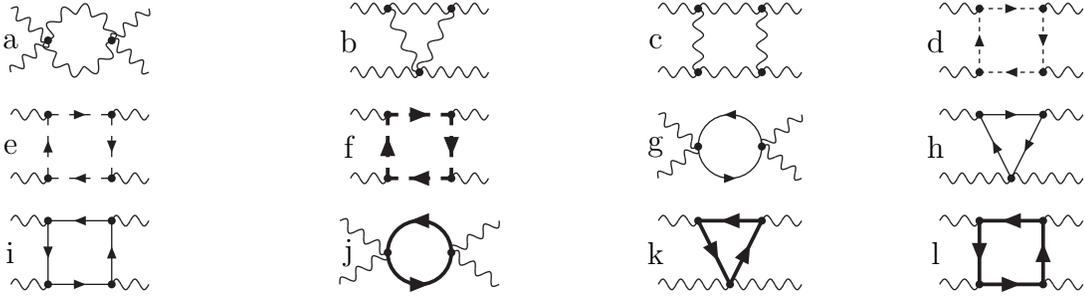

The diagrams which have UV divergent contributions to the three- and
four-point gauge field functions are shown in Figures~\ref{3Ph} and \ref
{4Ph}
respectively.
%%%%%%%%%%%%%%%%%%%%%%%%%%%%%%%%%%%%%%%%%%%%%%%%%
As in the case of the two-point gauge field function we generalize
our consideration to
an arbitrary number of the matter fields. The result is
\begin{eqnarray}
\label{A3}
S_{1A^3}&=&\frac{1}{(4\pi)^2}\frac{\frac{i}{3}g^3}{d-4}
  \int d^dx\, \mbox{tr}\left(
     \partial^\mu A^\nu\star{}\left[A_\mu,A_\nu\right]
     \right)\times
\\&&{}
\nonumber
  \begin{array}{lcccccr}
  {}&a+b+c&d&e&f&g&{}
  \\
  \times
  \left[\right.&N(9\alpha-17)&+8n_f&+16Nn_F&+2n_b&+4Nn_B&\left.\right]
  \end{array}
\end{eqnarray}
for the counterterm proportional to $A^3$ and
\begin{eqnarray}
\label{A4}
S_{1A^4}&=&\frac{1}{(4\pi)^2}\frac{\frac{1}{3}g^4}{d-4}
  \int d^dx\, \mbox{tr}\left(
  \,A_\mu\star{} A_\nu \star{}\left[A^\mu,A^\nu\right]
  \right)\times
\\&&{}
\nonumber
    \begin{array}{lcccccr}
    {} & a+b+c+d & e & f & g+h+i & j+k+l & {}
    \\
    \times[ & N(6\alpha-4) & +4n_f & +8Nn_F & +n_b & +2Nn_B & ]
    \end{array}
\end{eqnarray}
for the four-point counterterm.
From (\ref{A3}) we get the renormalization of the gauge field
coupling constant (we keep all the renormalized coupling constants
to be dimensionless) which is the same as in commutative $SU(N)$
gauge theory with the same matter field content
\begin{eqnarray}
\label{g}
\stackrel{\circ}{g}&=&\mu^{\frac{4-d}{2}}Z_gg\qquad
     Z_g=1+\frac{1}{(4\pi)^2}\frac{\frac{1}{3}g^2}{d-4}
     \left[N(22-2n_B-8n_F)-n_b-4n_f\right],
\end{eqnarray}
where $\mu$ is an arbitrary parameter with dimension of mass.
As far as the counterterm (\ref{A4}) is concerned it is absorbed by the 
renormalization
of the gauge field (\ref{A}) and the gauge coupling constant (\ref{g}).

Note here that the structure of the non-planar contributions to the
three-point 1PI gauge field function has the form
\begin{eqnarray}
&&
\label{NPAAA1}
\int_{k_1k_2k_3}\tilde\delta(k_1+k_2+k_3)\
  \mbox{tr}\Bigl(\tilde A_\alpha(k_1)\tilde A_\beta(k_2)\Bigr)\
  \mbox{tr}\tilde A_\gamma(k_3)
 \int_k f^{\alpha\beta\gamma}_1(k,k_1,k_2,k_3)
\\&+&
\int_{k_1k_2k_3}\tilde\delta(k_1+k_2+k_3)\
  \mbox{tr}\tilde A_\alpha(k_1)\
  \mbox{tr}\tilde A_\beta(k_2)\
  \mbox{tr}\tilde A_\gamma(k_3)
 \int_k f^{\alpha\beta\gamma}_2(k,k_1,k_2,k_3)
\label{NPAAA2}
\end{eqnarray}
($f^{\alpha\beta\gamma}$ are some functions) and as in the case of the
two-point 1PI gauge field function there is no pure $SU(N)$ contribution 
here
(\ref{NPAAA1},\ref{NPAAA2}) due to tracelessness of generators $T_a$ of 
$SU(N)$ group.

Also note that the quantity
\begin{eqnarray}
\label{gA}
Z_gZ_A&=&1+\frac{1}{(4\pi)^2}\frac{g^2N}{d-4}(\alpha+3)
\end{eqnarray}
is independent of the presence of the matter fields (as well as in the
commutative $SU(N)$ gauge field theory).

%%%%%%%%%%%%%%%%%%%%%%%%%%%%%%%%%%%%%%%%%%%%%%%

\subsection{1PI functions with ghost field external lines}

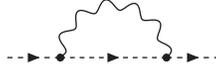
\begin{figure}[tbhp]
\begin{picture}(455,40)(0,0)
%--------------------------------
\Vertex(205,0){1.5}
\Vertex(245,0){1.5}
\DashArrowLine(185,0)(205,0){2}
\DashArrowLine(205,0)(245,0){2}
\DashArrowLine(245,0)(265,0){2}
\PhotonArc(225,0)(20,0,180){2}{6}
\SetWidth{0.5}
%--------------------------------
\end{picture}
\caption{\protect
\parbox[t]{10cm}{\protect\centering{Diagrams contributing to the two-
point function
              of  ghost field}}
        }\label{2Gh}
\end{figure}

There is only one diagram contributing in the two-point 1PI
ghost field function. It is shown in Figure \ref{2Gh}
and the
counterterm which results from it has the form
\begin{eqnarray}
\label{2C}
S_{1C^2}&=&
     \frac{1}{(4\pi)^2}\frac{g^2N}{d-4}(3-\alpha)\int d^dx\
        \mbox{tr}\Bigl( \bar{C}\partial^2C \Bigr).
\end{eqnarray}
The renormalization of the ghost fields is easily found from
(\ref{2C})
\begin{eqnarray}
\label{C}
\stackrel{\circ}{C}&=&Z_C\ C \qquad
  Z_C=1+\frac{1}{(4\pi)^2}\frac{g^2N}{d-4}\frac{\alpha-3}{2}.
\end{eqnarray}
The number of all the fields in the adjoint representation
(including the ghost fields) is greater by one in comparison
with its number in commutative $SU(N)$ gauge field theory due to
the
existence one more field corresponding $U(1)$ generator of
$U(N)$ group.
The renormalization of $SU(N)$ part of the ghost
fields in the noncommutative case
is the same as in the commutative $SU(N)$ gauge field theory.

\begin{figure}[tbhp]
\begin{picture}(455,40)(0,0)
\Vertex(105,0){1.5}
\Vertex(135,0){1.5}
\Vertex(120,25){1.5}
\DashArrowLine(85,0)(105,0){2}
\DashArrowLine(105,0)(135,0){2}
\DashArrowLine(135,0)(155,0){2}
\Photon(105,0)(120,25){2}{3}
\Photon(135,0)(120,25){2}{3}
\Photon(120,40)(120,25){2}{2}
%----------------------------------
\Vertex(330,0){1.5}
\Vertex(300,0){1.5}
\Vertex(315,25){1.5}
\DashArrowLine(280,0)(300,0){2}
\DashArrowLine(300,0)(315,25){2}
\DashArrowLine(315,25)(330,0){2}
\DashArrowLine(330,0)(350,0){2}
\Photon(315,25)(315,40){2}{2}
\Photon(330,0)(300,0){2}{4}
\end{picture}
\caption{\protect
\parbox[t]{10cm}{\protect\centering{Diagrams contributing to the three-
point function
              of ghost field coupling
              to gauge field}}
        }\label{3Gh}
\end{figure}
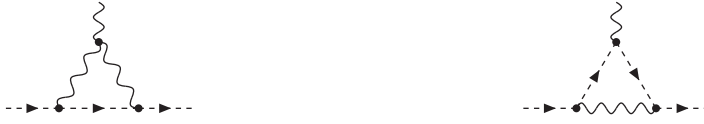

Turning to the three-point function of the ghost field coupled to
the gauge field. The relevant diagrams are shown in Figure
\ref{3Gh}
%%%%%%%%%%%%%%%%%%%%%%%%%%%%%%%%%%%%%%%%%%%%%%%%%%
which result in the following counterterm
\begin{eqnarray*}
S_{1C^2A}&=&
\frac{1}{(4\pi)^2}\frac{2i\alpha{}g^3N}{d-4}\int d^dx\
  \mbox{tr}\Bigl( \bar{C}\star\partial^\mu [A_\mu,C]\Bigr) .
\end{eqnarray*}
This counterterm is absorbed by the renormalization of the
fields and the gauge coupling constant (\ref{gA},\ref{C})
and does not violate multiplicative renormalizability of the
theory and $U(N)$ gauge invariance at the one-loop level.

%%%%%%%%%%%%%%%%%%%%%%%%%%%%%%%%%%%%%%%%%%%%%%%

\subsection{1PI functions with gauge field and fermion external lines}
\label{34}
\begin{figure}[htbp]
\begin{picture}(455,30)(0,0)
\Vertex(105,0){1.5}
\Vertex(155,0){1.5}
\PhotonArc(130,0)(25,0,180){3}{7}
\DashArrowLine(85,0)(105,0){5}
\DashArrowLine(105,0)(155,0){5}
\DashArrowLine(155,0)(175,0){5}
\Text(95,10)[]{a}
%----------------------------------
\Vertex(350,0){1.5}
\Vertex(300,0){1.5}
\ArrowArcn(325,0)(25,180,0)
\SetWidth{1.4}
\DashArrowLine(300,0)(350,0){5}
\SetWidth{0.5}
\DashArrowLine(350,0)(371,0){5}
\DashArrowLine(279,0)(300,0){5}
\Text(290,10)[]{b}
\end{picture}
\caption{\protect
\parbox[t]{10cm}{\protect\centering{Diagrams contributing to the two-
point function
              of a fermion field in the fundamental representation}}
        }\label{2f}
\end{figure}
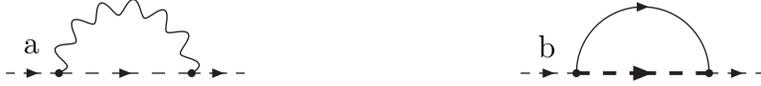

Let us first consider 1PI two-point function of the fermion in the 
fundamental
represntation. There are only two diagrams contributing to this function
(Fig.\ref{2f}),
%%%%%%%%%%%%%%%%%%%%%%%%%%%%%%%%%%%%%%%%%%%%%%%%%
which have no non-planar contributions (and as a consequence there is no
UV/IR mixing here).
They lead to the counterterms
\begin{eqnarray*}
S_{1\psi^2}&=&\frac{1}{(4\pi)^2}\frac{N}{d-4} \left(2\alpha g^2-|h|^2
\right)
        \int d^dx\,\bar{\psi}i\gamma^\mu\partial_\mu\psi
\\&{}&
       +\frac{1}{(4\pi)^2}\frac{-2N}{d-4}\left(m_1(\alpha+3)g^2+M_1|h|^2
\right)
        \int d^dx\,\bar{\psi}\psi.
\end{eqnarray*}
which have the same structure as in commutative $SU(N)$ theory but differ 
by
the numerical coefficients.

Now we try to generalize the result to the case of an arbitrary number
of the matter fields.
The relevant part of the classical
action should have the following form
\begin{eqnarray}
&& \int d^dx 
   \Bigl(
   \sum_{A=1}^{n_f}
   {\bar\psi}_{A}\star i\gamma^\mu D_\mu\psi_A
   -
   \sum_{A,A'=1}^{n_f}
   {\bar\psi}_{A}m_{1AA'}\psi_{A'}\Bigr)
\nonumber
\\
&-&\sum_{A=1}^{n_f}\sum_{B=1}^{n_F}\sum_{C=1}^{n_b}
  \int d^dx
    \left(
 h_{ABC}\,{\bar\psi}_{A}\star{}\Psi_{B}\star{}\phi_{C}
+h^{*}_{ABC}\,\phi^{+}_{C}\star{}{\bar\Psi}_{B}\star{}\psi_{A}
    \right).
\label{ABC}
\end{eqnarray}
In general case the mass matrix $m_{1AA'}$ is a constant
hermitian matrix.

Let us briefly describe the general structure of
renormalization of the field ${\bar\psi}_{A}$ and its mass
matrix. After calculating the one-loop counterterms
the relevant part of the classical action plus the couterterms
have the form
\begin{eqnarray}
\int d^dx \sum_{A,A'=1}^{n_f}
          \left\{
          \bar{\psi}_{A}\left[\delta_{AA'}+\frac{1}{d-4}E_{AA'}\right]
          i\gamma^\mu\partial_\mu\psi_{A'}
          -
          \bar{\psi}_{A}\left[m_{1AA'}+\frac{1}{d-4}M_{AA'}
          \right]\psi_{A'}
          \right\}.
\label{genf}
\end{eqnarray}
Here $E_{AA'}$ and $M_{AA'}$ are some constant hermitian
matrices $E_{AA'}^*=E_{A'A}$, $M_{AA'}^*=M_{A'A}$ generated by
the divergences.
From the first term of (\ref{genf}) we get the renormalization of the
field
\begin{eqnarray*}
\stackrel{\circ}{\psi}_A&=&\sum_{A'=1}^{n_f}\left(
       \delta_{AA'}+\frac{1}{d-4}\frac{1}{2}E_{AA'}
       \right)
       \psi_{A'}.
%\label{psiA}
\end{eqnarray*}
Having expressed
the renormalized field $\psi_A$ from the bare one
$\stackrel{\circ}{\psi}_A$ we substitute $\psi_A$ to
(\ref{genf}) and get the mass term in the form
\begin{eqnarray*}
-\sum_{A,A',A''=1}^{n_f}\int d^dx
    \stackrel{\circ}{\bar{\psi}}_A
  \left[
     m_{1AA'}
     +
     \frac{1}{d-4}
     \Big(M_{AA'}
     -\frac{1}{2}E_{AA''}m_{1A''A'}
     -\frac{1}{2}m_{1AA''}E_{A''A'}
     \Big)
  \right]
  \stackrel{\circ}{\psi}_{A'}.
%\label{2term}
\end{eqnarray*}
From this expression we see that the renormalization of the mass
matrix looks like
\begin{eqnarray*}
\stackrel{\circ}{m}_{1AA'}=m_{1AA'}
 +\frac{1}{d-4}\left[M_{AA'}
  -\frac{1}{2}\sum_{A''=1}^{n_f}
   \Big(
        E_{AA''}m_{1A''A'}
        +
        m_{1AA''}E_{A''A'}
   \Big)
   \right].
\end{eqnarray*}
Note that if we assume that the renormalized mass matrix is diagonal
$m_{1AA'}=m_{1A}\delta_{AA'}$, then in general case the bare
mass matrix $\stackrel{\circ}{m}_{1AA'}$ can't be diagonal
since neither $M_{AA'}$ nor $E_{AA'}$ must be diagonal. It
should also be noted that  the above general structure of the
renormalizations of the field and the mass matrix is independent
of whether the theory is noncommutative or not.

Further we will discuss mainly the features associated with
relationship between renormalizations of
commutative and noncommutative theories. To avoid the unessential
complications and tedious relations and
understand how mass renormalization is
organized in noncommutative models we consider a special
situation when the bare mass matrix is diagonal
$\stackrel{\circ}{m}_{1AA'}=\stackrel{\circ}{m}_{1A}\delta_{AA'}$
and the corresponding renormalized matrix is also diagonal
$m_{1AA'}=m_{1A}\delta_{AA'}$.

Taking into account the above assumption we get the relevant
part of the classical action
plus the counterterms in the form (\ref{genf}) where
\begin{eqnarray}
\nonumber
E_{AA'}&=&\frac{N}{(4\pi)^2}\left(
          2\alpha{}g^2\delta_{AA'}
          -\sum_{B=1}^{n_F}\sum_{C=1}^{n_b}h_{ABC}h^*_{A'BC}
          \right),
\\
M_{AA'}&=&\frac{2N}{(4\pi)^2}\left(
          m_{1A}g^2(3+\alpha)\delta_{AA'}
          +\sum_{B=1}^{n_F}\sum_{C=1}^{n_b}M_{1B}h_{ABC}h^*_{A'BC}
          \right).
\label{EM}
\end{eqnarray}
From (\ref{EM}) we see that $E_{AA'}$ and $M_{AA'}$ are not
diagonal and if we demand the bare mass matrix
$\stackrel{\circ}{m}_{1AA'}$ to be diagonal then
we must impose the following restrictions on the parameters of
the theory
\begin{eqnarray*}
\sum_{B=1}^{n_F}\sum_{C=1}^{n_b}
   \Big(
   4M_{1B}+m_{1A}+m_{1A'}
   \Big)
   h_{ABC}h^*_{A'BC}
=C_{A}\delta_{AA'},
\end{eqnarray*}
with $C_A$ being some quantities.
Since all the diagrams contributing to this 1PI functions are
planar, these restrictions are the same as in the corresponding
commutative theory.
Therefore the interaction (\ref{ABC}) should be adapted in the
proper way or be discarded completely.

In the case of a single fermion field in
the fundamental representation we have
\begin{eqnarray}
\label{psi}
\stackrel{\circ}{\psi}&=&Z_\psi\psi\qquad
   Z_\psi=1+\frac{1}{(4\pi)^2}\frac{N}{d-4}
   \left(\alpha g^2-\frac{1}{2}\sum_{B=1}^{n_F}\sum_{C=1}^{n_b}|h_{BC}|^2
\right),
\\
\label{m}
\stackrel{\circ}{m_1}&=&m_1+\frac{1}{(4\pi)^2}\frac{N}{d-4}
   \left[6g^2m_1+\sum_{B=1}^{n_F}\sum_{C=1}^{n_b}(m_1+2M_{1B})|h_{BC}|^2
   \right].
\end{eqnarray}
Note here that the renormalization of the fermionic field in the
fundamental representation and its mass
have the same structure
as in the commutative theory and the same numerical
coefficients due to the absence of nonplanar diagrams contributing
to the 1PI function in this case. It is interesting to point out
that although the interaction Lagrangians in commutative and
noncommutative theories differ, the renormalization relations
(\ref{psi},\ref{m}) under the above restrictions, turned out to
be the same in both cases.

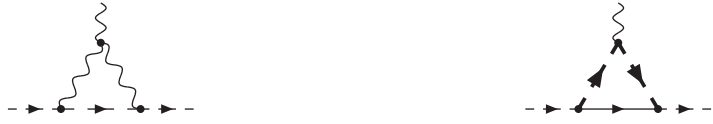
\begin{figure}[htbp]
\begin{picture}(455,40)(0,0)
\Vertex(105,0){1.5}
\Vertex(135,0){1.5}
\Vertex(120,25){1.5}
\DashArrowLine(85,0)(105,0){5}
\DashArrowLine(105,0)(135,0){5}
\DashArrowLine(135,0)(155,0){5}
\Photon(105,0)(120,25){2}{3}
\Photon(135,0)(120,25){2}{3}
\Photon(120,40)(120,25){2}{2}
%----------------------------------
\Vertex(330,0){1.5}
\Vertex(300,0){1.5}
\Vertex(315,25){1.5}
\ArrowLine(300,0)(330,0)
\DashArrowLine(280,0)(300,0){5}
\DashArrowLine(330,0)(350,0){5}
\SetWidth{1.4}
\DashArrowLine(300,0)(315,25){5}
\DashArrowLine(315,25)(330,0){5}
\SetWidth{0.5}
\Photon(315,25)(315,40){2}{2}
\end{picture}
\caption{\protect
\parbox[t]{10cm}{\protect\centering{Diagrams contributing to the three-
point function
              of a fermion field in the fundamental representation 
coupling
              to gauge field}}
        }\label{3f}
\end{figure}

Let us examine the fermion-gauge field vertex.
The relevant diagrams are shown in Figure \ref{3f}.
%%%%%%%%%%%%%%%%%%%%%%%%%%%%%%%%%%%%%%%%%%%%%%%%%%
Their contributions in the case of a single fermion field
in the fundamental representation is
\begin{eqnarray*}
S_{1\psi^2A}&=&
\frac{1}{(4\pi)^2}\frac{gN}{d-4}\left(3g^2(\alpha+1)
   -\sum_{B=1}^{n_F}\sum_{C=1}^{n_b}|h_{BC}|^2\right)
  \int d^dx\,\bar{\psi}\star{}\gamma^\mu A_\mu\star{}\psi.
\end{eqnarray*}
Since the renormalization of the fields $\psi$ and $A_\mu$ and the gauge
coupling constant $g$ have already been done, in the general case this
counterterm may break the multiplicative renormalizability of the theory.
But this does not happen
due to the preservation of $U(N)$ gauge invariance at the
one-loop level
and it is absorbed by
the renormalization
of the spinor and
the gauge fields and the renormalization of
the gauge coupling constant (\ref{gA},\ref{psi}).

Note here that nonplanar contributions to this three-point 1PI
function are independent of $SU(N)$ part of the gauge fields.

Similar situation arises for the fermion field in the adjoint 
representation.
The diagrams are shown in Figure~\ref{2F} and \ref{3F}.
\begin{figure}[htbp]
\begin{picture}(455,30)(0,0)
\Vertex(105,0){1.5}
\Vertex(155,0){1.5}
\PhotonArc(130,0)(25,0,180){3}{7}
\SetWidth{1.4}
\DashArrowLine(85,0)(105,0){5}
\DashArrowLine(105,0)(155,0){5}
\DashArrowLine(155,0)(175,0){5}
\SetWidth{0.5}
\Text(95,10)[]{a}
%----------------------------------
\Vertex(350,0){1.5}
\Vertex(300,0){1.5}
\ArrowArc(325,0)(25,0,180)
\DashArrowLine(300,0)(350,0){5}
\SetWidth{1.4}
\DashArrowLine(350,0)(371,0){5}
\DashArrowLine(279,0)(300,0){5}
\SetWidth{0.5}
\Text(290,10)[]{b}
\end{picture}
\caption{\protect
\parbox[t]{10cm}{\protect\centering{Diagrams contributing to the two-
point function
              of a fermion field in the adjoint representation}}
        }\label{2F}
\end{figure}
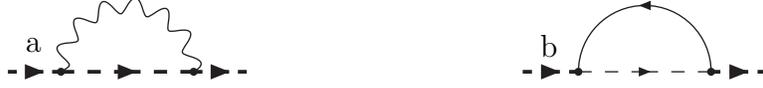
%%%%%%%%%%%%%%%%%%%%%%%%%%%%%%%%%%%%%%%%%%%%%%%%%%%%%%%%%%
\begin{figure}[htbp]
\begin{picture}(455,40)(0,0)
\Vertex(40,0){1.5}
\Vertex(80,0){1.5}
\Vertex(65,25){1.5}
\SetWidth{1.4}
\DashArrowLine(30,0)(50,0){5}
\DashArrowLine(50,0)(80,0){5}
\DashArrowLine(80,0)(100,0){5}
\SetWidth{0.5}
\Photon(50,0)(65,25){2}{3}
\Photon(80,0)(65,25){2}{3}
\Photon(65,40)(65,25){2}{2}
%----------------------------------
\Vertex(160,0){1.5}
\Vertex(190,0){1.5}
\Vertex(175,25){1.5}
\Photon(175,25)(175,40){2}{2}
\Photon(160,0)(190,0){2}{3}
\SetWidth{1.4}
\DashArrowLine(140,0)(160,0){5}
\DashArrowLine(160,0)(175,25){5}
\DashArrowLine(175,25)(190,0){5}
\DashArrowLine(190,0)(210,0){5}
\SetWidth{0.5}
%----------------------------------
\Vertex(270,0){1.5}
\Vertex(300,0){1.5}
\Vertex(285,25){1.5}
\Photon(285,25)(285,40){2}{2}
\DashArrowLine(270,0)(300,0){5}
\ArrowLine(300,0)(285,25)
\ArrowLine(285,25)(270,0)
\SetWidth{1.4}
\DashArrowLine(250,0)(270,0){5}
\DashArrowLine(300,0)(320,0){5}
\SetWidth{0.5}
%----------------------------------
\Vertex(410,0){1.5}
\Vertex(380,0){1.5}
\Vertex(395,25){1.5}
\ArrowLine(410,0)(380,0)
\SetWidth{1.4}
\DashArrowLine(360,0)(380,0){5}
\DashArrowLine(410,0)(430,0){5}
\SetWidth{0.5}
\DashArrowLine(380,0)(395,25){5}
\DashArrowLine(395,25)(410,0){5}
\Photon(395,25)(395,40){2}{2}
\end{picture}
\caption{\protect
\parbox[t]{10cm}{\protect\centering{Diagrams contributing to the three-
point function
              of a fermion field in the adjoit representation coupling to
              gauge field}}
        }\label{3F}
\end{figure}
The counterterms coming from these diagrams in case of one fermion
field in the adjoint representation are
\begin{eqnarray}
\label{Psi2}
&&\phantom{+}
\frac{1}{(4\pi)^2}\frac{1}{d-4}
  \left(4\alpha g^2N-\sum_{A=1}^{n_f}\sum_{C=1}^{n_b}|h_{AC}|^2\right)
  \int d^dx\, \mbox{tr}\left(
  \,\bar{\Psi}i\gamma^\mu\partial_\mu\Psi
  \right)
\\&{}&
\label{PsiM}
+\frac{-1}{(4\pi)^2}\frac{1}{d-4}
  \left(4(\alpha+3)g^2NM_1
   +2\sum_{A=1}^{n_f}\sum_{C=1}^{n_b}m_{1A}|h_{AC}|^2\right)
  \int d^dx \, \mbox{tr}\left(
  \,\bar{\Psi}\Psi   \right)
\\&{}&
\label{PsiA}
+\frac{1}{(4\pi)^2}\frac{g}{d-4}
  \left((3+5\alpha)g^2N-\sum_{A=1}^{n_f}\sum_{C=1}^{n_b}|h_{AC}|^2\right)
  \int d^dx\, \mbox{tr}\left(
  \,\bar{\Psi}\star{}\gamma^\mu\left[A_\mu,\Psi\right]
  \right).
\end{eqnarray}
From (\ref{Psi2}) we have the renormalization of the field $\Psi$
\begin{eqnarray}
\label{Psi0}
\stackrel{\circ}{\Psi}&=&Z_\Psi\Psi\qquad
   Z_\Psi =1+\frac{1}{(4\pi)^2}\frac{1}{d-4}
   \left(2\alpha g^2N
    -\frac{1}{2}\sum_{A=1}^{n_f}\sum_{C=1}^{n_b}|h_{AC}|^2\right),
\end{eqnarray}
and from (\ref{PsiM}) we have the renormalization of the mass $M_1$
\begin{eqnarray}
\label{M}
\stackrel{\circ}{M_1}&=&M_1+\frac{1}{(4\pi)^2}\frac{1}{d-4}
   \left[12g^2NM_1
    +\sum_{A=1}^{n_f}\sum_{C=1}^{n_b}\left(2m_{1A}+M_{1}\right)|h_{AC}|^2
    \right].
\end{eqnarray}
Thus, we see, the fermion masses (\ref{m},\ref{M})
are mixed with each other only in the presence of the
boson-fermion interaction (\ref{ABC}).
It should be also noted that renormalization of the $SU(N)$ part of
the fermionic field
in the adjoint representation and its mass
are the same as in the commutative case.
Counterterm (\ref{PsiA}), as it may easily be checked, is absorbed by the
renormalization (\ref{gA},\ref{Psi0}).

As far as the nonplanar contributions to these 1PI functions are
concerned their structure is similar to (\ref{NPAA}) and
(\ref{NPAAA1},\ref{NPAAA2}) for the cases of the two- and
three-point functions respectively.
And, as a consequence, the nonplanar contribution to the
two-point function depends on the $U(1)$ parts of the fields only
and nonplanar contribution to the three-point function has no
pure $SU(N)$ field dependence.

%%%%%%%%%%%%%%%%%%%%%%%%%%%%%%%%%%%%%%%%%%%%%%%%%%%%%%%%%%%%%%%%%%%%%%%
%%%%%%%%%%%%%%%%%%%%%%%%%%%%%%%%%%%%%%%%%%%%%%%%%%%%%%%%%%%%%%%%%%%%%%%
%%%%%%%%%%%%%%%%%%%%%%%%%%%%%%%%%%%%%%%%%%%%%%%%%%%%%%%%%%%%%%%%%%%%%%%

\subsection{1PI functions with gauge field and boson external lines}

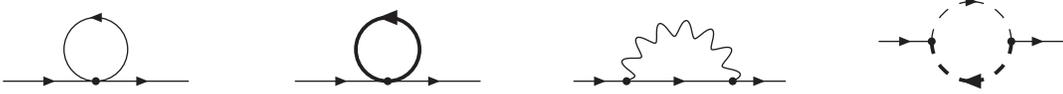
\begin{figure}[thbp]
\begin{picture}(455,40)(0,0)
\ArrowArc(65,12)(12,-90,270)
\ArrowLine(30,0)(65,0)
\ArrowLine(65,0)(100,0)
\Vertex(65,0){1.5}
%----------------------------------
\SetWidth{1.4}
\ArrowArc(175,12)(12,-90,270)
\SetWidth{0.5}
\ArrowLine(140,0)(175,0)
\ArrowLine(175,0)(210,0)
\Vertex(175,0){1.5}
%----------------------------------
\Vertex(265,0){1.5}
\Vertex(305,0){1.5}
\ArrowLine(245,0)(265,0)
\ArrowLine(265,0)(305,0)
\ArrowLine(305,0)(325,0)
\PhotonArc(285,0)(20,0,180){3}{7}
%----------------------------------
\Vertex(410,15){1.5}
\Vertex(380,15){1.5}
\ArrowLine(360,15)(380,15)
\ArrowLine(410,15)(430,15)
\DashArrowArcn(395,15)(15,180,360){5}
\SetWidth{1.4}
\DashArrowArcn(395,15)(15,0,180){5}
\SetWidth{0.5}
\end{picture}
\caption{\protect
\parbox[t]{10cm}{\protect\centering{Diagrams contributing to the two-
point function
              of a boson field in the fundamental representation}}
        }\label{2b}
\end{figure}

For
the boson field in the fundamental representation the diagrams 
corresponding
to its 1PI two-point function
are
shown in Figure~\ref{2b},
which result in the counterterms
\begin{eqnarray*}
&&\phantom{+}
 \frac{1}{(4\pi)^2}\frac{2N}{d-4}\left(g^2(\alpha-3)+2|h|^2\right)
  \int d^dx\,\partial^\mu\phi^+\partial_\mu\phi
\\&{}&
+\frac{1}{(4\pi)^2}\frac{1}{d-4}
  \left(\frac{\lambda_1}{3!}m_2^2(N+1)+2N(f_a+f_b)M_2^2-2\alpha g^2Nm_2^2
  \right.
  \\&{}&
  \left.
  \phantom{\frac{1}{(4\pi)^2}\frac{1}{d-4}}
  -8N|h|^2(m_1^2+m_1M_1+M_1^2)\right)
  \int d^dx\,\phi^+\phi.
\end{eqnarray*}
This 1PI function as well as 1PI function of the fermionic field
in the fundamental representation has no nonplanar contribution.

Let us try to generalize these counterterms to the case of an arbitrary
number of the matter fields. First of all we must write down the relevant
part of the classical action. It has the following form
\begin{eqnarray}
&&\nonumber
   \int d^dx\ \Bigl\{
   D_\mu\phi^+_CD^\mu\phi_C-\phi^+_Cm^2_{2CC'}\phi_{C'}
   -\frac{\lambda_{1C_1C_2C_3C_4}}{4!}\
   \phi^+_{C_1}\star\phi_{C_2}\star\phi^+_{C_3}\star\phi_{C_4}
\\&&\label{theSS}
 -f_{aC_1D_2D_1C_2}\phi^+_{C_1}\star\Phi_{D_2}\star\Phi^+_{D_1}\star\phi_
{C_2}
 -f_{bC_1D_1D_2C_2}\phi^+_{C_1}\star\Phi^+_{D_1}\star\Phi_{D_2}\star\phi_
{C_2}
\\&&\nonumber
 -h_{ABC}\,{\bar\psi}_{A}\star{}\Psi_{B}\star{}\phi_{C}
 -h^{*}_{ABC}\,\phi^{+}_{C}\star{}{\bar\Psi}_{B}\star{}\psi_{A}
   \Bigl\}.
\end{eqnarray}
Hereafter summing over repeated indices {\sc a,b,c,d} is assumed.
Indices {\sc a} run from 1 to $n_f$, {\sc b} run from 1 to $n_F$,
{\sc c} run from 1 to $n_b$, {\sc d} run from 1 to $n_B$.
In expression (\ref{theSS}) $f_{aC_1D_2D_1C_2}$ and $f_{bC_1D_1D_2C_2}$ 
are
real constants and $\lambda_1$ has the symmetry
\begin{eqnarray*}
  \lambda_{1C_1C_2C_3C_4}
  =\lambda_{1C_3C_4C_1C_2}
  =\lambda^*_{1C_2C_1C_4C_3}
  =\lambda^*_{1C_4C_3C_2C_1},
\end{eqnarray*}
which follows from the reality condition of the action and properties of 
the
star-product.

The structure of renormalization of the field $\phi_C$ and
its mass matrix is similar to that in case of the fermionic
field in the fundamental representation which was described
earlier. It is also independent of whether the theory is
noncommutative or not. To simplify the calculations we assume
like in section \ref{34} that both bare and renormalized mass
matrices are diagonal
$\stackrel{\circ}{m}{}^2_{2CC'}=\stackrel{\circ}{m}{}^2_{2C}\delta_{CC'}$,
$m^2_{2CC'}=m^2_{2C}\delta_{CC'}$.
Then the relevant part of the classical action plus the
counterterms have the form
\begin{eqnarray*}
\int d^dx\ \partial^\mu\phi^+_{C_1}\partial_\mu\phi_{C_2}
  \Bigl( \delta_{C_1C_2}+\frac{1}{d-4}E_{C_1C_2}\Bigr)
  -\phi^+_{C_1}\phi_{C_2}
  \Bigl(m^2_{2C_1}\delta_{C_1C_2}+\frac{1}{d-4}M_{C_1C_2}\Bigr).
%\label{phiA}
\end{eqnarray*}
Here
\begin{eqnarray*}
E_{C_1C_2}&=&\frac{2N}{(4\pi)^2}
     \Bigl(
     g^2(\alpha-3)\delta_{C_1C_2}+2h^*_{ABC_1}h_{ABC_2}
     \Bigr),
\\
M_{C_1C_2}&=&\frac{1}{(4\pi)^2}
     \Bigl(
     2\alpha g^2Nm^2_{2C_1}\delta_{C_1C_2}
     +8Nh^*_{ABC_1}h_{ABC_2}(m^2_{1A}+m_{1A}M_{1B}+M^2_{1B})
\\&&
     -2NM^2_{2D}(f_{aC_1DDC_2}+f_{bC_1DDC_2})
     -\frac{m^2_{2C}}{3!}(N\lambda_{1CCC_1C_2}+\lambda_{1C_1CCC_2})
     \Bigr).
\end{eqnarray*}
Doing similar calculations as in case of the field $\psi_A$
we at first get the renormalization of the bosonic field in the
fundamental representation
\begin{eqnarray*}
\stackrel{\circ}{\phi}_C&=&\left(
  \delta_{CC'}+\frac{1}{d-4}\frac{1}{2}E_{CC'}
  \right)\phi_{C'}.
%\label{renphiA}
\end{eqnarray*}
and then the renormalization of its mass matrix
\begin{eqnarray*}
\stackrel{\circ}{m}{}^2_{2CC'}
=
 m^2_{2C}\delta_{CC'}+\frac{1}{d-4}\left[
 M_{CC'}-\frac{1}{2}m^2_{2C}E_{CC'}
 -\frac{1}{2}E_{CC'}m^2_{2C'}
 \right].
\end{eqnarray*}
If we demand the bare mass matrix
$\stackrel{\circ}{m}{}^2_{2CC'}$ to be also diagonal
we must impose the following restrictions on the parameters of
the theory
\begin{eqnarray*}
&&
2Nh^*_{ABC_1}h_{ABC_2}
  \left(
  4m^2_{1A}+4m_{1A}M_{1B}+4M^2_{1B}-m^2_{2C_1}-m^2_{2C_2}
  \right)
\\&&
 -2NM^2_{2D}
  \left(
        f_{aC_1DDC_2}+f_{bC_1DDC_2}
  \right)
 -\frac{m^2_{2C}}{3!}\left(N\lambda_{1CCC_1C_2}+\lambda_{1C_1CCC_2}\right)
 =A_C\delta_{CC'},
\end{eqnarray*}
with $A_C$ being some quantities and summing over indices
{\sc a}, {\sc b}, {\sc c} and {\sc d} is assumed.
This
situation is completely analogous to that in the commutative
theory.
Since there are no nonplanar diagrams contributing to the 1PI
function under consideration then the same restriction arise in
the corresponding commutative theory.

In the following we shall not discuss the generalization of the
theory to an
arbitrary number of the matter fields. In the case of one field of each 
type
we have the renormalization of $\phi$ and $m_2$
\begin{eqnarray}
\label{phi}
\stackrel{\circ}{\phi}&=&Z_\phi\phi\qquad
   Z_\phi=1+\frac{1}{(4\pi)^2}\frac{N}{d-4}
   \left[(\alpha-3)g^2+2|h|^2\right],
\\
\nonumber
\stackrel{\circ}{m}{}_2^2&=&m_2^2+\frac{1}{(4\pi)^2}\frac{1}{d-4}
   \left[4N|h|^2(2m_1^2+2m_1M_1+2M_1^2-m_2^2)+6g^2Nm_2^2
   \vphantom{\frac{\lambda_1}{3!}}\right.
   \\ &{}&
   \phantom{m_2^2+\frac{1}{(4\pi)^2}\frac{1}{d-4}}\left.
   -2N(f_a+f_b)M_2^2-\frac{\lambda_1}{3!}(N+1)m_2^2\right].
\end{eqnarray}
The renormalization of the bosonic field in the
fundamental representation and its mass are the same as in the
corresponding commutative theory due to the absence of the
nonplanar diagrams contributing to the relevant 1PI function.

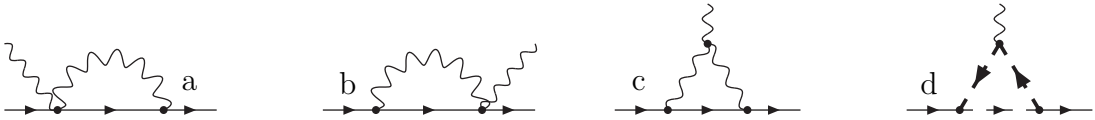
\begin{figure}[htbp]
\begin{picture}(455,40)(0,0)
\Vertex(40,0){1.5}
\Vertex(80,0){1.5}
\ArrowLine(20,0)(40,0)
\ArrowLine(40,0)(80,0)
\ArrowLine(80,0)(100,0)
\PhotonArc(60,0)(20,0,180){3}{7}
\Photon(40,0)(20,25){2}{4}
\Text(90,10)[]{a}
%----------------------------------
\Vertex(160,0){1.5}
\Vertex(200,0){1.5}
\PhotonArc(180,0)(20,0,180){3}{7}
\ArrowLine(140,0)(160,0)
\ArrowLine(160,0)(200,0)
\ArrowLine(200,0)(220,0)
\Photon(200,0)(220,25){2}{4}
\Text(150,10)[]{b}
%----------------------------------
\Vertex(270,0){1.5}
\Vertex(300,0){1.5}
\Vertex(285,25){1.5}
\Photon(285,25)(285,40){2}{2}
\ArrowLine(270,0)(300,0)
\Photon(300,0)(285,25){2}{3}
\Photon(285,25)(270,0){2}{3}
\ArrowLine(250,0)(270,0)
\ArrowLine(300,0)(320,0)
\Text(260,10)[]{c}
%----------------------------------
\Vertex(410,0){1.5}
\Vertex(380,0){1.5}
\Vertex(395,25){1.5}
\DashArrowLine(380,0)(410,0){5}
\SetWidth{1.4}
\DashArrowLine(395,25)(380,0){5}
\DashArrowLine(410,0)(395,25){5}
\SetWidth{0.5}
\ArrowLine(360,0)(380,0)
\ArrowLine(410,0)(430,0)
\Photon(395,25)(395,40){2}{2}
\Text(370,10)[]{d}
\end{picture}
\caption{\protect
\parbox[t]{10cm}{\protect\centering{Diagrams contributing to the three-
point function
              of a boson field in the fundamental representation coupling 
to
              gauge field}}
        }\label{3b}
\end{figure}

Since we have already renormalized the gauge field coupling,
the gauge field and the bosonic field in the fundamental representation
we should check that these renormalization relations absorb the 
divergencies of
the three- and four-point 1PI functions.
The divergent
diagrams corresponding to three-point function are shown in Figure~\ref
{3b}.
Summing up the contributions of these diagrams we find the counterterm
\begin{eqnarray*}
&{}&
\frac{1}{(4\pi)^2}\frac{igN}{d-4}
  \begin{array}[b]{rcccl}
  {}&a+b&c&d&{}\\
  \Bigl(&-3g^2&+3\alpha{}g^2&+4|h|^2&\Bigr)
  \end{array}
  \int d^dx\,\left[\phi^+\star{}A_\mu\star{}\partial^\mu\phi
  -\partial^\mu\phi^+\star{}A_\mu\star{}\phi\right],
\end{eqnarray*}
which is absorbed by the renormalization of the fields and the
gauge coupling constant (\ref{gA},\ref{phi}).

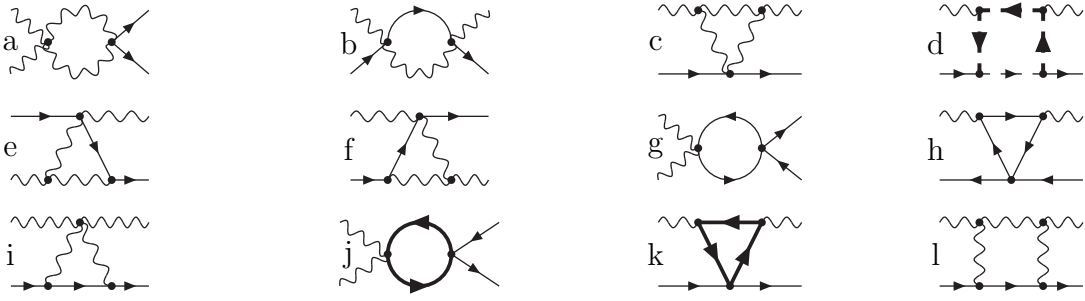
\begin{figure}[htbp]
\begin{picture}(455,110)(0,0)
%-----------------------------------------
\Vertex(50,92){1.5}
\Vertex(74,92){1.5}
\PhotonArc(62,92)(12,0,360){2}{10}
\Photon(50,92)(36,104){2}{3}
\Photon(50,92)(36,80){2}{3}
\ArrowLine(74,92)(88,104)
\ArrowLine(74,92)(88,80)
\Text(36,92)[]{\mbox{}a}
%-----------------------------------------
\Vertex(202,92){1.5}
\Vertex(178,92){1.5}
\PhotonArc(190,92)(12,180,360){2}{5.5}
\ArrowArcn(190,92)(12,180,0)
\Photon(202,92)(216,104){2}{3}
\Photon(178,92)(164,104){2}{3}
\ArrowLine(202,92)(216,80)
\ArrowLine(164,80)(178,92)
\Text(164,92)[]{\mbox{}b}
%-----------------------------------------
\Vertex(295,104){1.5}
\Vertex(319,104){1.5}
\Vertex(307,80){1.5}
\Photon(295,104)(319,104){2}{3}
\Photon(319,104)(307,80){2}{3}
\Photon(307,80)(295,104){2}{3}
\Photon(280,104)(295,104){2}{2}
\Photon(319,104)(334,104){2}{2}
\ArrowLine(280,80)(307,80)
\ArrowLine(307,80)(334,80)
\Text(280,92)[]{\mbox{}c}
%-----------------------------------------
\Vertex(401,104){1.5}
\Vertex(425,104){1.5}
\Vertex(401,80){1.5}
\Vertex(425,80){1.5}
\SetWidth{1.5}
\DashArrowLine(425,104)(401,104){5}
\DashArrowLine(425,80)(425,104){5}
\DashArrowLine(401,104)(401,80){5}
\SetWidth{0.5}
\DashArrowLine(401,80)(425,80){5}
\Photon(401,104)(386,104){2}{2}
\Photon(425,104)(440,104){2}{2}
\ArrowLine(425,80)(440,80)
\ArrowLine(386,80)(401,80)
\Text(386,92)[]{\mbox{}d}
%-----------------------------------------
%-----------------------------------------
\Vertex(50,40){1.5}
\Vertex(62,64){1.5}
\Vertex(74,40){1.5}
\ArrowLine(36,64)(62,64)
\ArrowLine(62,64)(74,40)
\ArrowLine(74,40)(88,40)
\Photon(50,40)(36,40){2}{2}
\Photon(50,40)(62,64){2}{3}
\Photon(50,40)(74,40){2}{3}
\Photon(62,64)(88,64){2}{3}
\Text(36,52)[]{\mbox{}e}
%-----------------------------------------
\Vertex(178,40){1.5}
\Vertex(202,40){1.5}
\Vertex(190,64){1.5}
\ArrowLine(164,40)(178,40)
\ArrowLine(178,40)(190,64)
\ArrowLine(190,64)(216,64)
\Photon(164,64)(190,64){2}{3}
\Photon(190,64)(202,40){2}{3}
\Photon(178,40)(202,40){2}{3}
\Photon(202,40)(216,40){2}{2}
\Text(164,52)[]{\mbox{}f}
%-----------------------------------------
\Vertex(295,52){1.5}
\Vertex(319,52){1.5}
\ArrowArc(307,52)(12,0,180)
\ArrowArc(307,52)(12,180,360)
\Photon(295,52)(280,64){2}{3}
\Photon(295,52)(280,40){2}{3}
\ArrowLine(334,40)(319,52)
\ArrowLine(319,52)(334,64)
\Text(280,52)[]{\mbox{}g}
%-----------------------------------------
\Vertex(401,64){1.5}
\Vertex(425,64){1.5}
\Vertex(413,40){1.5}
\ArrowLine(440,40)(413,40)
\ArrowLine(413,40)(386,40)
\Photon(401,64)(386,64){2}{2}
\Photon(425,64)(440,64){2}{2}
\ArrowLine(401,64)(425,64)
\ArrowLine(425,64)(413,40)
\ArrowLine(413,40)(401,64)
\Text(386,52)[]{\mbox{}h}
%-----------------------------------------
%-----------------------------------------
\Vertex(74,0){1.5}
\Vertex(50,0){1.5}
\Vertex(62,24){1.5}
\ArrowLine(36,0)(50,0)
\ArrowLine(50,0)(74,0)
\ArrowLine(74,0)(88,0)
\Photon(36,24)(62,24){2}{4}
\Photon(62,24)(88,24){2}{4}
\Photon(62,24)(74,0){2}{3}
\Photon(62,24)(50,0){2}{3}
\Text(36,12)[]{\mbox{}i}
%-----------------------------------------
\SetWidth{1.4}
\ArrowArc(190,12)(12,0,180)
\ArrowArc(190,12)(12,180,360)
\SetWidth{0.5}
\Vertex(178,12){1.5}
\Vertex(202,12){1.5}
\ArrowLine(202,12)(220,0)
\ArrowLine(220,24)(202,12)
\Photon(178,12)(160,24){2}{3}
\Photon(178,12)(160,0){2}{3}
\Text(164,12)[]{\mbox{}j}
%-----------------------------------------
\Vertex(307,0){1.5}
\Vertex(319,24){1.5}
\Vertex(295,24){1.5}
\ArrowLine(280,0)(307,0)
\ArrowLine(307,0)(334,0)
\Photon(280,24)(295,24){2}{2}
\Photon(334,24)(319,24){2}{2}
\SetWidth{1.4}
\ArrowLine(307,0)(319,24)
\ArrowLine(319,24)(295,24)
\ArrowLine(295,24)(307,0)
\SetWidth{0.5}
\Text(280,12)[]{\mbox{}k}
%-----------------------------------------
\Vertex(425,0){1.5}
\Vertex(425,24){1.5}
\Vertex(401,0){1.5}
\Vertex(401,24){1.5}
\Photon(425,24)(440,24){2}{2}
\ArrowLine(425,0)(440,0)
\ArrowLine(386,0)(401,0)
\Photon(386,24)(401,24){2}{2}
\Photon(425,0)(425,24){2}{3}
\Photon(425,24)(401,24){2}{3}
\Photon(401,24)(401,0){2}{3}
\ArrowLine(401,0)(425,0)
\Text(386,12)[]{\mbox{}l}
\end{picture}
\caption{\protect
\parbox[t]{10cm}{\protect\centering{Diagrams contributing to the four-
point
         function of a boson field in the fundamental representation 
coupling
         to gauge field}}
        }\label{4b}
\end{figure}

The diagrams contributing to the four-point function are shown in
Figure~\ref{4b}.\footnote{Diagrams which are not shown in Figure~\ref{4b} 
are non-planar.}
Diagrams g and h, j and k, i and l, cancel each other. The others give the
following contribution to the counterterm
\begin{eqnarray*}
&&
\frac{1}{(4\pi)^2}\frac{g^2N}{d-4}
  \int d^dx\,\phi^+\star{}A_\mu\star{} A^\mu \star{}\phi\times
\\&&{}\times
  \begin{array}[b]{rcccccl}
  {}&a&b&c&d&e+f(=e)&{}\\
  (&-\frac{3}{2}g^2(3+\alpha^2)&-\frac{1}{2}g^2(3+\alpha)&
   +\frac{3}{2}g^2(4+\alpha+\alpha^2)&
   +4|h|^2&
   +3\alpha{}g^2&)\,,
  \end{array}
\end{eqnarray*}
which is also absorbed by the renormalization of the fields and the gauge
coupling constant (\ref{gA},\ref{phi}).

\begin{figure}[htbp]
\begin{picture}(455,40)(0,0)
\Vertex(100,0){1.5}
\SetWidth{1.5}
\ArrowLine(65,0)(100,0)
\ArrowLine(100,0)(135,0)
\SetWidth{0.5}
\ArrowArc(100,15)(15,-90,270)
%--------------------------------
\Vertex(225,0){1.5}
\SetWidth{1.4}
\ArrowLine(190,0)(225,0)
\ArrowLine(225,0)(260,0)
\ArrowArc(225,15)(15,-90,270)
\SetWidth{0.5}
%--------------------------------
\Vertex(330,0){1.5}
\Vertex(370,0){1.5}
\PhotonArc(350,0)(20,0,180){2}{6}
\SetWidth{1.5}
\ArrowLine(310,0)(330,0)
\ArrowLine(330,0)(370,0)
\ArrowLine(370,0)(390,0)
\SetWidth{0.5}
\end{picture}
\caption{\protect
\parbox[t]{10cm}{\protect\centering{Diagrams contributing to the two-
point function
              of a boson field in the adjoint representation}}
        }\label{2B}
\end{figure}
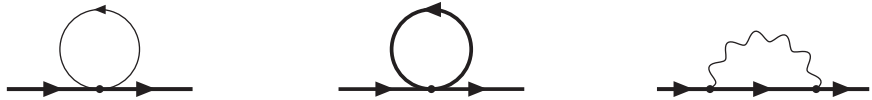

For the case of the boson field in the adjoint representation we have 
diagrams in
Figure~\ref{2B}
for the two-point 1PI function. The resulting counterterm and 
renormalization
relations
of the field $\Phi$ and the mass $M_2$ are
\begin{eqnarray*}
\\&{}&
\phantom{+}
 \frac{1}{(4\pi)^2}\frac{4g^2N}{d-4}(\alpha-3)
  \int d^dx \,\mbox{tr}\left(
  \,\partial^\mu\Phi^+\partial_\mu\Phi
  \right)
\\&{}&
+\frac{1}{(4\pi)^2}\frac{1}{d-4}
  \left[2m_2^2(f_a+f_b)
     +NM_2^2\left(\frac{1}{3!}(2\lambda_{2a}+\lambda_{2b})
     -4\alpha g^2\right)\right]
  \int d^dx \,\mbox{tr}\left(
  \,\Phi^+\Phi
  \right),
\end{eqnarray*}
\begin{eqnarray}
\stackrel{\circ}{\Phi}&=&Z_\Phi\Phi\qquad
   Z_\Phi=1+\frac{1}{(4\pi)^2}\frac{2g^2N}{d-4}(\alpha-3),
\label{Phi}
\\
\nonumber
\stackrel{\circ}{M_2^2}&=&M_2^2+\frac{1}{(4\pi)^2}\frac{1}{d-4}
   \left[12g^2NM_2^2-\frac{1}{3!}N(2\lambda_{2a}+\lambda_{2b})M_2^2
   -2(f_a+f_b)m_2^2\right].
\end{eqnarray}
The renormalization of $SU(N)$ part of the bosonic field in the
adjoint representation
is the same as in the
commutative $SU(N)$ gauge field theory with the same matter
field content.

\begin{figure}[htbp]
\begin{picture}(455,40)(0,0)
\Vertex(40,0){1.5}
\Vertex(80,0){1.5}
\SetWidth{1.4}
\ArrowLine(20,0)(40,0)
\ArrowLine(40,0)(80,0)
\ArrowLine(80,0)(100,0)
\SetWidth{0.5}
\PhotonArc(60,0)(20,0,180){3}{7}
\Photon(40,0)(20,25){2}{4}
\Text(90,10)[]{a}
%----------------------------------
\Vertex(160,0){1.5}
\Vertex(200,0){1.5}
\PhotonArc(180,0)(20,0,180){3}{7}
\SetWidth{1.4}
\ArrowLine(140,0)(160,0)
\ArrowLine(160,0)(200,0)
\ArrowLine(200,0)(220,0)
\SetWidth{0.5}
\Photon(200,0)(220,25){2}{4}
\Text(150,10)[]{b}
%----------------------------------
\Vertex(270,0){1.5}
\Vertex(300,0){1.5}
\Vertex(285,25){1.5}
\Photon(285,25)(285,40){2}{2}
\Photon(300,0)(285,25){2}{3}
\Photon(285,25)(270,0){2}{3}
\SetWidth{1.4}
\ArrowLine(270,0)(300,0)
\ArrowLine(250,0)(270,0)
\ArrowLine(300,0)(320,0)
\SetWidth{0.5}
\Text(260,10)[]{c}
%----------------------------------
\Vertex(410,0){1.5}
\Vertex(380,0){1.5}
\Vertex(395,25){1.5}
\SetWidth{1.4}
\ArrowLine(360,0)(380,0)
\ArrowLine(380,0)(395,25)
\ArrowLine(395,25)(410,0)
\ArrowLine(410,0)(430,0)
\SetWidth{0.5}
\Photon(395,25)(395,40){2}{2}
\Photon(380,0)(410,0){2}{3}
\Text(370,10)[]{d}
\end{picture}
\caption{\protect
\parbox[t]{10cm}{\protect\centering{Diagrams contributing to the three-
point function
              of a boson field in the adjoint representation}}
        }\label{3B}
\end{figure}
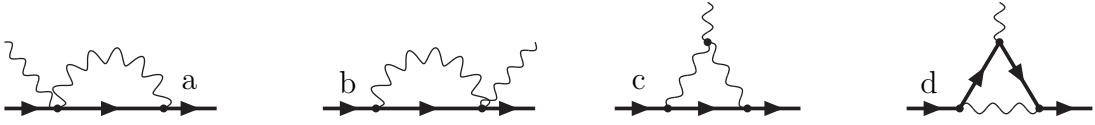

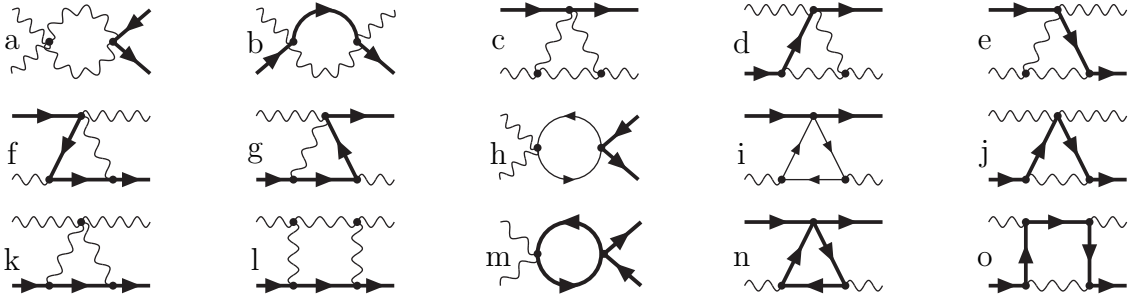
\begin{figure}[htbp]
\begin{picture}(455,110)(0,0)
%---------------------------------
\Vertex(29,92){1.5}
\Vertex(53,92){1.5}
\SetWidth{1.4}
\ArrowLine(67,104)(53,92)
\ArrowLine(53,92)(67,80)
\SetWidth{0.5}
\Photon(29,92)(15,104){2}{3}
\Photon(29,92)(15,80){2}{3}
\PhotonArc(41,92)(12,0,360){2}{10}
\Text(15,92)[]{a}
%---------------------------------
\Vertex(121,92){1.5}
\Vertex(145,92){1.5}
\SetWidth{1.4}
\ArrowLine(107,80)(121,92)
\ArrowLine(145,92)(159,80)
\ArrowArcn(133,92)(12,180,0)
\SetWidth{0.5}
\Photon(107,104)(121,92){2}{3}
\Photon(145,92)(159,104){2}{3}
\PhotonArc(133,92)(12,180,0){2}{5.5}
\Text(107,92)[]{b}
%---------------------------------
\Vertex(225,104){1.5}
\Vertex(213,80){1.5}
\Vertex(237,80){1.5}
\SetWidth{1.4}
\ArrowLine(199,104)(225,104)
\ArrowLine(225,104)(251,104)
\SetWidth{0.5}
\Photon(199,80)(213,80){2}{2}
\Photon(213,80)(225,104){2}{3}
\Photon(213,80)(237,80){2}{3}
\Photon(237,80)(225,104){2}{3}
\Photon(237,80)(251,80){2}{2}
\Text(199,92)[]{c}
%---------------------------------
\Vertex(317,104){1.5}
\Vertex(305,80){1.5}
\Vertex(329,80){1.5}
\SetWidth{1.4}
\ArrowLine(291,80)(305,80)
\ArrowLine(305,80)(317,104)
\ArrowLine(317,104)(343,104)
\SetWidth{0.5}
\Photon(291,104)(317,104){2}{4}
\Photon(329,80)(317,104){2}{3}
\Photon(329,80)(305,80){2}{3}
\Photon(329,80)(343,80){2}{2}
\Text(291,92)[]{d}
%---------------------------------
\Vertex(409,104){1.5}
\Vertex(397,80){1.5}
\Vertex(421,80){1.5}
\SetWidth{1.4}
\ArrowLine(383,104)(409,104)
\ArrowLine(409,104)(421,80)
\ArrowLine(421,80)(435,80)
\SetWidth{0.5}
\Photon(383,80)(397,80){2}{2}
\Photon(397,80)(409,104){2}{3}
\Photon(397,80)(421,80){2}{3}
\Photon(409,104)(435,104){2}{4}
\Text(383,92)[]{e}
%---------------------------------
%---------------------------------
\Vertex(29,40){1.5}
\Vertex(53,40){1.5}
\Vertex(41,64){1.5}
\SetWidth{1.4}
\ArrowLine(15,64)(41,64)
\ArrowLine(41,64)(29,40)
\ArrowLine(29,40)(53,40)
\ArrowLine(53,40)(67,40)
\SetWidth{0.5}
\Photon(15,40)(29,40){2}{2}
\Photon(41,64)(67,64){2}{4}
\Photon(41,64)(53,40){2}{3}
\Text(15,50)[]{f}
%---------------------------------
\Vertex(133,64){1.5}
\Vertex(121,40){1.5}
\Vertex(145,40){1.5}
\SetWidth{1.4}
\ArrowLine(107,40)(121,40)
\ArrowLine(121,40)(145,40)
\ArrowLine(145,40)(133,64)
\ArrowLine(133,64)(159,64)
\SetWidth{0.5}
\Photon(107,64)(133,64){2}{4}
\Photon(133,64)(121,40){2}{3}
\Photon(145,40)(159,40){2}{2}
\Text(107,50)[]{g}
%---------------------------------
\Vertex(237,52){1.5}
\Vertex(213,52){1.5}
\ArrowArc(225,52)(12,180,0)
\ArrowArc(225,52)(12,0,180)
\SetWidth{1.4}
\ArrowLine(251,64)(237,52)
\ArrowLine(237,52)(251,40)
\SetWidth{0.5}
\Photon(199,40)(213,52){2}{3}
\Photon(199,64)(213,52){2}{3}
\Text(199,50)[]{h}
%---------------------------------
\Vertex(305,40){1.5}
\Vertex(329,40){1.5}
\Vertex(317,64){1.5}
\Photon(343,40)(329,40){2}{2}
\Photon(291,40)(305,40){2}{2}
\ArrowLine(305,40)(317,64)
\ArrowLine(317,64)(329,40)
\ArrowLine(329,40)(305,40)
\SetWidth{1.4}
\ArrowLine(291,64)(317,64)
\ArrowLine(317,64)(343,64)
\SetWidth{0.5}
\Text(291,50)[]{i}
%---------------------------------
\Vertex(397,40){1.5}
\Vertex(421,40){1.5}
\Vertex(409,64){1.5}
\Photon(383,64)(409,64){2}{4}
\Photon(435,64)(409,64){2}{4}
\Photon(397,40)(421,40){2}{3}
\SetWidth{1.4}
\ArrowLine(383,40)(397,40)
\ArrowLine(397,40)(409,64)
\ArrowLine(409,64)(421,40)
\ArrowLine(421,40)(435,40)
\SetWidth{0.5}
\Text(383,50)[]{j}
%---------------------------------
%---------------------------------
\Vertex(29,0){1.5}
\Vertex(53,0){1.5}
\Vertex(41,24){1.5}
\SetWidth{1.4}
\ArrowLine(15,0)(29,0)
\ArrowLine(29,0)(53,0)
\ArrowLine(53,0)(67,0)
\SetWidth{0.5}
\Photon(41,24)(15,24){2}{4}
\Photon(41,24)(67,24){2}{4}
\Photon(41,24)(29,0){2}{3}
\Photon(41,24)(53,0){2}{3}
\Text(15,10)[]{k}
%---------------------------------
\Vertex(121,0){1.5}
\Vertex(145,0){1.5}
\Vertex(121,24){1.5}
\Vertex(145,24){1.5}
\SetWidth{1.4}
\ArrowLine(107,0)(121,0)
\ArrowLine(121,0)(145,0)
\ArrowLine(145,0)(159,0)
\SetWidth{0.5}
\Photon(107,24)(121,24){2}{2}
\Photon(121,24)(145,24){2}{3}
\Photon(145,24)(159,24){2}{2}
\Photon(121,24)(121,0){2}{3}
\Photon(145,24)(145,0){2}{3}
\Text(107,10)[]{l}
%---------------------------------
\Vertex(213,12){1.5}
\Vertex(238,12){1.5}
\Photon(199,24)(213,12){2}{2}
\Photon(199,0)(213,12){2}{2}
\SetWidth{1.5}
\ArrowArc(225,12)(12,180,0)
\ArrowArc(225,12)(12,0,180)
\ArrowLine(238,12)(252,24)
\ArrowLine(252,0)(238,12)
\SetWidth{0.5}
\Text(199,10)[]{m}
%---------------------------------
\Vertex(317,24){1.5}
\Vertex(305,0){1.5}
\Vertex(329,0){1.5}
\Photon(291,0)(305,0){2}{2}
\Photon(329,0)(343,0){2}{2}
\SetWidth{1.4}
\ArrowLine(291,24)(317,24)
\ArrowLine(317,24)(343,24)
\ArrowLine(317,24)(329,0)
\ArrowLine(329,0)(305,0)
\ArrowLine(305,0)(317,24)
\SetWidth{0.5}
\Text(291,10)[]{n}
%---------------------------------
\Vertex(397,0){1.5}
\Vertex(397,24){1.5}
\Vertex(421,24){1.5}
\Vertex(421,0){1.5}
\Photon(383,24)(397,24){2}{2}
\Photon(421,24)(435,24){2}{2}
\Photon(397,0)(421,0){2}{3}
\SetWidth{1.4}
\ArrowLine(383,0)(397,0)
\ArrowLine(397,0)(397,24)
\ArrowLine(397,24)(421,24)
\ArrowLine(421,24)(421,0)
\ArrowLine(421,0)(435,0)
\SetWidth{0.5}
\Text(383,10)[]{o}
%---------------------------------
\end{picture}
\caption{\protect
\parbox[t]{10cm}{\protect\centering{Diagrams contributing to the four-
point function
              of a boson field in the adjoint representation}}
        }\label{4B}
\end{figure}

Since the three- and four-point 1PI functions depend only on the gauge
coupling constant, the gauge field and the bosonic field in the adjoint
representation for which renormalization has already done we
need to check that the
divergences of these 1PI functions are absorbed by the fields and the
gauge coupling constant renormalization.
Corresponding diagrams are shown in
Figure~\ref{3B}
for the three-point function and Figure~\ref{4B} for the four-point 
function.
%---------------------------------------------
Calculating these diagrams one obtains
\begin{eqnarray*}
S_{1\Phi^2A}
&=&
\frac{1}{(4\pi)^2}\frac{ig^3N}{d-4}
  \int d^dx \, \mbox{tr}
  \,\left(\left[\Phi^+,A_\mu\right]\star{}\partial^\mu\Phi
               -\partial^\mu\Phi^+\star{}\left[A_\mu,\Phi\right]\right)
  \times
\\ && {} \times
    \begin{array}[b]{rcccl}
    {} & a+b & c      & d       &{}\\ {}
    [  &  -9 &+2\alpha &+3\alpha &  ]
    \end{array}
\\
S_{1\Phi^2A^2}
&=&
  \frac{1}{(4\pi)^2}\frac{6g^4N}{d-4}
  (1-\alpha)
  \int d^dx\, \mbox{tr}\left(
  \,\bigl[A^\mu,\Phi^+\bigr]\star{}\bigl[A_\mu,\Phi\bigr]
  \right)
\\
 (a) &=&\phantom{+}
  \frac{1}{(4\pi)^2}\frac{g^4N}{d-4}
  \frac{9+3\alpha^2}{2}
  \int d^dx\, \mbox{tr}\left(
  \,\bigl[A^\mu,\Phi^+\bigr]\star{}\bigl[A_\mu,\Phi\bigr]
    -2\Phi^+\star{}A_\mu\star{}\Phi{}\star{}A^\mu
  \right)
\\
 (b) &{}&\mbox{}+
  \frac{1}{(4\pi)^2}\frac{g^4N}{d-4}
  \frac{3+\alpha}{2}
  \int d^dx\, \mbox{tr}\left(
  \,5\bigl[A^\mu,\Phi^+\bigr]\star{}\bigl[A_\mu,\Phi\bigr]
    -2\Phi^+\star{}A_\mu\star{}\Phi{}\star{}A^\mu
  \right)
\\
 (c) &{}&\mbox{}+
  \frac{1}{(4\pi)^2}\frac{g^4N}{d-4}
  \frac{-12-3\alpha-3\alpha^2}{2}
  \int d^dx\, \mbox{tr}\left(
  \,\bigl[A^\mu,\Phi^+\bigr]\star{}\bigl[A_\mu,\Phi\bigr]
  \right.
  \\&{}&\hspace{8cm}
  \left.
  \mbox{}
    -2\Phi^+\star{}A_\mu\star{}\Phi{}\star{}A^\mu
  \right)
\\
 (d+e(=d)) &{}&\mbox{}+
  \frac{1}{(4\pi)^2}\frac{g^4N}{d-4}
  \bigl(-3\alpha\bigr)
  \int d^dx\, \mbox{tr}\left(
  \,\bigl[A^\mu,\Phi^+\bigr]\star{}\bigl[A_\mu,\Phi\bigr]
    +2\Phi^+\star{}A_\mu\star{}\Phi{}\star{}A^\mu
  \right)
\\
 (f+g(=f)) &{}&\mbox{}+
  \frac{1}{(4\pi)^2}\frac{g^4N}{d-4}
  \bigl(-4\alpha\bigr)
  \int d^dx\, \mbox{tr}\left(
  \,\bigl[A^\mu,\Phi^+\bigr]\star{}\bigl[A_\mu,\Phi\bigr]
    -\Phi^+\star{}A_\mu\star{}\star{}\Phi{}\star{}A^\mu
  \right)
\end{eqnarray*}
For the four-point function contributions of h and i, j
and o, k and l, m and n diagrams cancel each other.
From these expressions
for the counterterms we see that
these counterterms are absorbed by the renormalization of the fields
and the gauge coupling constant (\ref{gA},\ref{Phi}).

The other 1PI functions can't destroy multiplicative renormalization of 
the
theory, since they can be absorbed by the renormalization of the coupling
constants of the matter fields for which renormalization has not
been done yet and we
may always absorb the divergences by their renormalization.
The rest of the counterterms are
\begin{eqnarray}
\nonumber
&&\phantom{+}
\frac{1}{(4\pi)^2}\frac{1}{d-4}
  \left[(N+1)\left(\frac{2\lambda_1}{4!}\right)^2
     +N(f_a^2+f_b^2)-\alpha g^2N\frac{\lambda_1}{3!}
  \right.
  \\&{}& \nonumber
  \left.
  \phantom{\frac{1}{(4\pi)^2}\frac{1}{d-4}}
     +3g^4N-4|h|^4N\right]
  \int d^dx\,\phi^+\star{}\phi\star{}\phi^+\star{}\phi
\\&{}& \label{L_A}
+\frac{1}{(4\pi)^2}\frac{1}{d-4}
  \left[6g^4N+f_a^2+f_b^2+\frac{2}{4!^2}N(4\lambda_{2a}^2+\lambda_{2b}^2)
  \right.
  \\&{}&
  \nonumber
  \left.
  \phantom{\frac{1}{(4\pi)^2}\frac{1}{d-4}}
     -\frac{1}{3}\alpha g^2N\lambda_{2a}\right]
  \int d^dx\, \mbox{tr}\left[
  \,\Phi^+\star{}\Phi\star{}\Phi^+\star{}\Phi
  \right]
\\&{}&  \label{L_B}
+\frac{1}{(4\pi)^2}\frac{1}{d-4}
  \left[6g^4N+2f_af_b+\frac{2}{4!^2}N\lambda_{2b}(4\lambda_{2a}+\lambda_
{2b})
     -\frac{1}{3}\alpha g^2N\lambda_{2b}
  \right.
  \\&{}&
  \nonumber
  \left.
  \phantom{\frac{1}{(4\pi)^2}\frac{1}{d-4}}
     +2N\left(\frac{|\lambda_3|}{3!}\right)^2\right]
  \int d^dx\, \mbox{tr}\left[
  \,\Phi^+\star{}\Phi^+\star{}\Phi\star{}\Phi
  \right]
\\&{}&
\nonumber
+\frac{1}{(4\pi)^2}\frac{1}{d-4}
  \left[6g^4N-6\alpha f_ag^2N+\frac{1}{3!}\lambda_1f_a+2f_a^2N
  \right.
  \\&{}&
  \nonumber
  \left.
  \phantom{\frac{1}{(4\pi)^2}\frac{1}{d-4}}
     +\frac{2}{4!}N(2f_a\lambda_{2a}+f_b\lambda_{2b}\right]
  \int d^dx\,\phi^+\star{}\Phi\star{}\Phi^+\star{}\phi
\\&{}&
\nonumber
+\frac{1}{(4\pi)^2}\frac{1}{d-4}
  \left[6g^4N-6\alpha f_bg^2N+\frac{1}{3!}\lambda_1f_b+2f_b^2N
  \right.
  \\&{}&
  \nonumber
  \left.
  \phantom{\frac{1}{(4\pi)^2}\frac{1}{d-4}}
     +\frac{2}{4!}N(2f_b\lambda_{2a}+f_a\lambda_{2b})\right]
  \int d^dx\,\phi^+\star{}\Phi^+\star{}\Phi\star{}\phi
\\&{}&
\nonumber
+\frac{1}{(4\pi)^2}\frac{2N}{d-4}\frac{\lambda_3}{3!}
  \left(\frac{\lambda_{2b}}{4!}-\alpha g^2\right)
  \int d^dx\, \mbox{tr}\left[
  \,\Phi\star{}\Phi\star{}\Phi\star{}\Phi
  \right]
\\&{}&
\nonumber
+\frac{1}{(4\pi)^2}\frac{2N}{d-4}\frac{\lambda_3^*}{3!}
  \left(\frac{\lambda_{2b}}{4!}-\alpha g^2\right)
  \int d^dx\, \mbox{tr}\left[
  \,\Phi^+\star{}\Phi^+\star{}\Phi^+\star{}\Phi^+
  \right]
\\&{}&
\nonumber
+\frac{1}{(4\pi)^2}\frac{g^2Nh}{d-4}(6+4\alpha)
   \int d^dx\,\bar{\psi}\star{}\Psi\star{}\phi
\\&{}&
\nonumber
+\frac{1}{(4\pi)^2}\frac{g^2Nh^*}{d-4}(6+4\alpha)
   \int d^dx\,\phi^+\star{}\bar{\Psi}\star{}\psi.
\end{eqnarray}
These counterterms lead to the following renormalization of
the coupling constants of the matter fields
\begin{eqnarray}
\nonumber
\mu^{d-4}\stackrel{\circ}{\frac{\lambda_1}{4!}}&=&\frac{\lambda_1}{4!}
   +\frac{1}{(4\pi)^2}\frac{1}{d-4}
   \left[4|h|^4N-3g^4N+\frac{1}{2}g^2N\lambda_1
   -\frac{1}{3}|h|^2N\lambda_1-N(f_a^2+f_b^2)
   \right.
   \\
   \nonumber
   &{}&
   \phantom{m_2^2+\frac{1}{(4\pi)^2}\frac{1}{d-4}}\left.
   -(N+1)\left(\frac{2\lambda_1}{4!}\right)^2\right],
\\
\nonumber
\mu^{d-4}\stackrel{\circ}{\frac{\lambda_{2a}}{4!}}&=&\frac{\lambda_{2a}}
{4!}
   +\frac{1}{(4\pi)^2}\frac{1}{d-4}
   \left[g^2N\lambda_{2a}
   -\frac{2}{(4!)^2}N(4\lambda^2_{2a}+\lambda^2_{2b})
   -f_a^2-f_b^2-6g^4N\right],
\\
\nonumber
\mu^{d-4}\stackrel{\circ}{\frac{\lambda_{2b}}{4!}}&=&\frac{\lambda_{2b}}
{4!}
   +\frac{1}{(4\pi)^2}\frac{1}{d-4}
   \left[g^2N\lambda_{2b}
   -\frac{2}{(4!)^2}N\lambda_{2b}(4\lambda_{2a}+\lambda_{2b})
   -2f_af_b\right.
   \\
   \nonumber
   &{}&
   \phantom{m_2^2+\frac{1}{(4\pi)^2}\frac{1}{d-4}}\left.
   -6g^4N-2N\left(\frac{|\lambda_3|}{3!}\right)^2\right],
\\
\nonumber
\mu^{d-4}\stackrel{\circ}{f_a}&=&f_a+\frac{1}{(4\pi)^2}\frac{1}{d-4}
   \left[18g^2Nf_a-6g^4N-\frac{\lambda_1}{3!}f_a-2f_a^2N
   -4|h|^2Nf_a\right.
   \\
   \nonumber
   &{}&
   \phantom{m_2^2+\frac{1}{(4\pi)^2}\frac{1}{d-4}}\left.
   -\frac{2}{4!}N(2f_a\lambda_{2a}+f_b\lambda_{2b})\right],
\\
\nonumber
\mu^{d-4}\stackrel{\circ}{f_b}&=&f_b+\frac{1}{(4\pi)^2}\frac{1}{d-4}
   \left[18g^2Nf_b-6g^4N-\frac{\lambda_1}{3!}f_b-2f_b^2N
   -4|h|^2Nf_b\right.
   \\
   \nonumber
   &{}&
   \phantom{m_2^2+\frac{1}{(4\pi)^2}\frac{1}{d-4}}\left.
   -\frac{2}{4!}N(2f_b\lambda_{2a}+f_a\lambda_{2b})\right],
\\
\label{h}
\stackrel{\circ}{h}&=&\mu^{\frac{4-d}{2}}Z_hh\qquad
   Z_h=1+\frac{1}{(4\pi)^2}\frac{1}{d-4}
   \left[\frac{1}{2}|h|^2(1-3N)-g^2N(3+8\alpha)\right],
\\
\label{lambda3}
\stackrel{\circ}{\lambda_3}&=&\mu^{4-d}Z_{\lambda_3}\lambda_{3}\qquad
   Z_{\lambda_3}=1+\frac{1}{(4\pi)^2}\frac{N}{d-4}
   \left[24g^2-\frac{1}{3}\lambda_{2b}\right].
\end{eqnarray}

As a result we see that the theory under consideration
is multiplicatively renormalizable in the one-loop
approximation.
If we consider $f_a$, $f_b$ and $\lambda_{2a}$, $\lambda_{2b}$ to be not
independent \cite{ABGKM,ABK,ABKR}
\begin{eqnarray*}
f_a\to{}fa_1, & \qquad f_b\to{}fb_1, & \qquad a_1+b_1=1,
\\
\lambda_{2a}\to\lambda_{2}a_2,
& \qquad
  \lambda_{2b}\to\lambda_{2}b_2,
& \qquad
  a_2+b_2=1,
\end{eqnarray*}
where $a$ and $b$ are real numbers (which are not renormalized),
then the theory will be
renormalizable if we put the following restrictions on
these numbers
\begin{eqnarray*}
a_1=b_1,\qquad & a_2=b_2, & \qquad \lambda_3=0.
\end{eqnarray*}

From the above formulae (\ref{h},\ref{lambda3}) we see that
if we would like to reduce the number of
interactions without breaking multiplicative renormalizability we could 
neglect only $h$ and $\lambda_3$ couplings.
It should be noted that these formulae of renormalization of the matter
fields coupling constants have never been written out in explicit form
in the literature.
For example in the works \cite{ABK,ABKR} only the structure of the
divergencies (\ref{L_A},\ref{L_B}) was discussed.

\section{Summary}
We have studied the one-loop renormalizability of the
general
noncommutative Yang-Mills field coupled to different kinds of matter 
fields
interacting among themselves.

Unlike all the previous works  we have included in the
action the scalar and the spinor matter fields both in the
fundamental and in the
adjoint representations.
The action also contains some new
terms describing interaction among the matter fields which have not been
considered previously in the context of the noncommutative field theories.
Naturally, inclusion of any new term in the action may
influence on renormalizability of the theory.
To prove the theory is one-loop multiplicatively renormalizable
we computed  all
counterterms needed to cancel the one-loop divergences of the effective 
action.
The formal structure of the counterterms of the noncommutative
theory is the same as the formal structure of the corresponding
counterterms of the commutative theory but pointwise
multiplication of the fields is replaced by the star product.
One more distinctive feature of the counterterms is the
numerical factors in the renormalization constants which differ
from the corresponding factors of the commutative theory due to
the appearance of nonplanar diagrams. The number of diagrams
contributing to a given 1PI function is the same as in a
commutative theory and its noncommutative counterpart. But some
of the diagrams of the noncommutative theory are nonplanar and so
have no UV divergences. This leads to difference of the
numerical factors in the renormalization constants. Since the
numerical factors are changed multiplicative
renormalizability of the theory may be destroyed but it does not
happen due to the preservation
of $U(N)$
gauge invariance of the model at the one-loop
level.

We have also shown that the result for pure gauge field 1PI
function may be generalized to the case of an arbitrary number of the 
matter
fields.
All results concerning the renormalization of the fields and
coupling constants agree with the previous results
in the literuture
and include them as a partial case.

Our calculations, in the framework of a general model confirm the
specific features of noncommutative field theories which were
found within the various simple models.
The number of UV divegent diagrams is
reduced due to the appearence of the nonplanar diagrams which
are considered to be UV finite.

On the whole, we have established the one-loop
multiplicative
renormalizability of general
noncommutative
Yang-Mills field model interacting with the matter fields.
At different values of its parameters, this model is reduced to
a number of various concrete models.
Therefore the results obtained here allow us to find the one-loop
counterterms for many concrete noncommutative field theories.

\section*{Acknowledgements}
The authors are grateful to
joint RFBR-DFG grant, project No.\ 02-02-04002,
DFG grant, project No.\ 436 RUS 113/669
and grant PD02-1{.}2-94 of Russian Ministry of Education
for partial support.
The work of I.L.B. was also supported in
part by the INTAS grant, project No.\ 00-00254.

\appendix
\section{Feynman rules}
%\footnote{${\tilde\delta}(k+p)=(2\pi)^d\delta(k+p)$}
\label{FR}

\underline{Propagators}
\begin{eqnarray*}
\begin{picture}(120,20)(0,6)
\SetOffset(25,10)
\Photon(20,0)(80,0){3}{5}
\Text(0,0)[]{$A_\mu{}^i_j(k)$}
\Text(105,0)[]{$A_\nu{}^{i'}_{j'}(k')$}
\end{picture}\hspace{30pt}
&=&
G_0(A_\mu{}^i_j(k);A_\nu{}^{i'}_{j'}(k'))
\\
&=&{\tilde\delta}(k+k')\delta^{i}_{j'}\delta^{i'}_{j}
  \left[
   -\frac{g_{\mu\nu}}{k^2}
      +(1-\alpha)\frac{k_{\mu}k_{\nu}}{k^4}
  \right]
\\
%------------------------------------------------------
\begin{picture}(120,20)(0,6)
%\SetWidth{1.8}
\SetOffset(25,10)
\DashArrowLine(20,0)(80,0){2}
\Text(0,0)[]{$\bar{C}^{r_1}_{s_1}(l_1)$}
\Text(105,0)[]{$C^{r_2}_{s_2}(l_2)$}
\end{picture}\hspace{30pt}
&=&
G_0(\bar{C}^{r_1}_{s_1}(l_1);C^{r_2}_{s_2}(l_2))
=\frac{{\tilde\delta}(l_1+l_2)\delta^{r_1}_{s_2}\delta^{r_2}_{s_1}}{l_1^2}
\\
%------------------------------------------------------
\begin{picture}(120,20)(0,6)
\SetWidth{1.8}
\SetOffset(25,10)
\DashArrowLine(20,0)(80,0){7}
\Text(-5,0)[]{$\bar{\Psi}^b{}^{k_1}_{l_1}(Q_1)$}
\Text(105,0)[]{$\Psi_a{}^{k_2}_{l_2}(Q_2)$}
\end{picture}\hspace{30pt}
&=&
G_0(\bar{\Psi}^b{}^{k_1}_{l_1}(Q_1);\Psi_a{}^{k_2}_{l_2}(Q_2))
\\
&=&
{\tilde\delta}(Q_1+Q_2)\delta^{k_1}_{l_2}\delta^{k_2}_{l_1}
 \frac{\gamma^\mu{}^b_aQ_{1\mu}+\delta^b_aM_1}{Q_1^2-M_1^2}
\\
%-------------------------------------------------------------
\begin{picture}(120,20)(0,6)
\SetOffset(25,10)
\DashArrowLine(20,0)(80,0){7}
\Text(0,0)[]{$\bar{\psi}^{bk}(q_1)$}
\Text(105,0)[]{$\psi_{al}(q_2)$}
\end{picture}\hspace{30pt}
&=&
G_0(\bar{\psi}^{bk}(q_1);\psi_{al}(q_2))
={\tilde\delta}(q_1+q_2)\delta^{k}_{l}
 \frac{\gamma^\mu{}^b_aq_{1\mu}+\delta^b_am_1}{q_1^2-m_1^2}
\\
%-------------------------------------------------------------
\begin{picture}(120,20)(0,6)
\SetWidth{1.8}
\SetOffset(25,10)
\ArrowLine(20,0)(80,0)
\Text(-5,0)[]{$\Phi^+{}^{m_1}_{n_1}(P_1)$}
\Text(105,0)[]{$\Phi{}^{m_2}_{n_2}(P_2)$}
\end{picture}\hspace{30pt}
&=&
G_0(\Phi^+{}^{m_1}_{n_1}(P_1);\Phi{}^{m_2}_{n_2}(P_2))
=\frac{{\tilde\delta}(P_1+P_2)\delta^{m_1}_{n_2}\delta^{m_2}_{n_1}}{P_1^2-
M_2^2}
\\
%---------------------------------------------------------------
\begin{picture}(120,20)(0,6)
\SetOffset(25,10)
\ArrowLine(20,0)(80,0)
\Text(-2,0)[]{$\phi^+{}^m(p_1)$}
\Text(105,0)[]{$\phi_n(p_2)$}
\end{picture}\hspace{30pt}
&=&
G_0(\phi^+{}^m(p_1);\phi_n(p_2))
=\frac{{\tilde\delta}(p_1+p_2)\delta^{m}_{n}}{p_1^2-m_2^2}
\\
{\tilde\delta}(k+p)&=&(2\pi)^d\delta(k+p)
\end{eqnarray*}

\noindent
\underline{Vertices}
\par\noindent
Here we have kept only terms which give contributions in the planar 
diagrams.
\begin{eqnarray*}
\begin{picture}(120,20)(0,6)
\SetOffset(25,10)
\Photon(20,0)(80,0){3}{5}
\Vertex(50,0){3}
\Text(0,0)[]{$A_\alpha{}^i_j(k)$}
\Text(105,0)[]{$A_\beta{}^{i'}_{j'}(k')$}
\end{picture}\hspace{30pt}
&=&V(A_\alpha{}^i_j(k);A_\beta{}^{i'}_{j'}(k'))
\\&&{}\hspace{-11em}
 =  g\int_{k_1}{\tilde\delta}(k+k'+k_1)
      \left[
       \delta^{j'}_{i}
       {\tilde A}_\mu{}^{j}_{i'}(k_1)
       e^{\frac{i}{2}k\theta{}k_1}
       -
       \delta^{j}_{i'}
       {\tilde A}_\mu{}^{j'}_{i}(k_1)
       e^{-\frac{i}{2}k\theta{}k_1}
       \right]
\\&&{}\hspace{-9em}
      \times
      \left[
      g_{\alpha\beta}(k^\mu-k'^\mu)
      +
      \delta^\mu_\alpha(k_{1\beta}-k_{\beta})
      +
      \delta^\mu_{\beta}(k'_\alpha-k_{1\alpha})
      \right]
\\&&{}\hspace{-11em}
+g^2\int_{k_1k_2}{\tilde\delta}(k+k'+k_1+k_2)
     \left\{
     e^{\frac{i}{2}k\theta(k_1+k_2)-\frac{i}{2}k_1\theta{}k_2}
     \delta^{j'}_{i}
     \right.
\\&&{}\hspace{-10em}
\times
\left[\,
     2{\tilde A}_\alpha{}^{m}_{i'}(k_1)
       {\tilde A}_\beta{}^{j}_{m}(k_2)
     -
     g_{\alpha\beta}{\tilde A}_\mu{}^{m}_{i'}(k_1)
       {\tilde A}^\mu{}^{j}_{m}(k_2)
     -
     {\tilde A}_\beta{}^{m}_{i'}(k_1)
       {\tilde A}_\alpha{}^{j}_{m}(k_2)
     \right]
\\&&{}\hspace{-10em}
     +
     e^{-\frac{i}{2}k\theta(k_1+k_2)-\frac{i}{2}k_1\theta{}k_2}
     \delta^{j}_{i'}
     \left[\,
     2{\tilde A}_\beta{}^{m}_{i}(k_1)
       {\tilde A}_\alpha{}^{j'}_{m}(k_2)
     \right.
\\&&{}\hspace{-9em}
     \left.
     -g_{\alpha\beta}{\tilde A}_\mu{}^{m}_{i}(k_1)
       {\tilde A}^\mu{}^{j'}_{m}(k_2)
     -
     {\tilde A}_\alpha{}^{m}_{i}(k_1)
       {\tilde A}_\beta{}^{j'}_{m}(k_2)
     \right]
\\&&{}\hspace{-11em}
+g^2\int_{p_1p_2}{\tilde\delta}(k+k'+p_1+p_2)g_{\alpha\beta}
      e^{\frac{i}{2}p_1\theta{}p_2}\times
\\&&{}\hspace{-11em}
 \qquad\left[\,
      \delta^{j'}_{i}
      {\tilde\phi}^{j}_*(p_1)
      {\tilde\phi}_{i'}(p_2)
      e^{\frac{i}{2}k\theta(p_1+p_2)}
   +
      \delta^{j}_{i'}
      {\tilde\phi}^{j'}_*(p_1)
      {\tilde\phi}_{i}(p_2)
      e^{-\frac{i}{2}k\theta(p_1+p_2)}
      \,\right]
\\&&{}\hspace{-11em}
+g^2\int_{P_1P_2}{\tilde\delta}(k+k'+P_1+P_2)g_{\alpha\beta}
   \left\{
   e^{\frac{i}{2}k\theta(P_1+P_2)}
   \delta^{j'}_{i}
   \right.
\\&&{}\hspace{-11em}
   \qquad
   \times
   \left[\,
     {\tilde\Phi}_{*}{}^{k}_{i'}(P_1)
                    {\tilde\Phi}{}^{j}_{k}(P_2)
                    e^{-\frac{i}{2}P_1\theta{}P_2}
     +
      {\tilde\Phi}_{*}{}^{j}_{k}(P_1)
                    {\tilde\Phi}{}^{k}_{i'}(P_2)
                    e^{\frac{i}{2}P_1\theta{}P_2}
    \,\right]
\\&&{}\hspace{-11em}
     \qquad
     +
     e^{-\frac{i}{2}k\theta(P_1+P_2)}
     \delta^{j}_{i'}
\\&&{}\hspace{-11em}
   \qquad
   \times
   \left[\,
    {\tilde\Phi}_{*}{}^{k}_{i}(P_1)
                    {\tilde\Phi}{}^{j'}_{k}(P_2)
                    e^{-\frac{i}{2}P_1\theta{}P_2}
    +
    {\tilde\Phi}_{*}{}^{j'}_{k}(P_1)
                    {\tilde\Phi}{}^{k}_{i}(P_2)
                    e^{\frac{i}{2}P_1\theta{}P_2}
   \,\right]
\\
%---------------------------------------------------
\begin{picture}(120,20)(0,6)
\SetOffset(25,10)
\Photon(20,0)(50,0){3}{2.5}
\DashArrowLine(50,0)(80,0){2}
\Vertex(50,0){3}
\Text(0,0)[]{$A_\mu{}^i_j(k)$}
\Text(105,0)[]{$\bar{C}^{r_1}_{s_1}(l_1)$}
\end{picture}\hspace{30pt}
&=&
V(A_\mu{}^i_j(k),\bar{C}^{r_1}_{s_1}(l_1))
=\frac{\delta_R}{\delta\tilde{\bar{C}}{}^{r_1}_{s_1}(l_1)}
 \frac{\delta{}V}{\delta\tilde{A}_\mu{}^i_j(k)}
\\&&\hspace{-10em}
 =g\int_{l_2}\tilde{\delta}(l_1+l_2+k)l_1^\mu
     \left[
     \delta^{s_1}_{i}\tilde{C}^j_{r_1}(l_2)
     e^{-\frac{i}{2}l_1\theta{}l_2}
     -
     \delta^j_{r_1}\tilde{C}^{s_1}_i(l_2)
     e^{\frac{i}{2}l_1\theta{}l_2}
     \right]
\\
%---------------------------------------------------
\begin{picture}(120,20)(0,6)
\SetOffset(25,10)
\Photon(20,0)(50,0){3}{2.5}
\DashArrowLine(80,0)(50,0){2}
\Vertex(50,0){3}
\Text(0,0)[]{$A_\mu{}^i_j(k)$}
\Text(105,0)[]{$C^{r_2}_{s_2}(l_2)$}
\end{picture}\hspace{30pt}
&=&
V(A_\mu{}^i_j(k),C^{r_2}_{s_2}(l_2))
\\&&\hspace{-10em}
 =g\int_{l_1}\tilde{\delta}(l_1+l_2+k)l_1^\mu
     \left[
     \delta^{s_2}_{i}\tilde{\bar{C}}{}^j_{r_2}(l_1)
     e^{\frac{i}{2}l_1\theta{}l_2}
     -
     \delta^j_{r_2}\tilde{\bar{C}}{}^{s_2}_i(l_1)
     e^{-\frac{i}{2}l_1\theta{}l_2}
     \right]
\\
%---------------------------------------------------
\begin{picture}(120,20)(0,6)
\SetOffset(25,10)
\Photon(20,0)(50,0){3}{2.5}
\DashArrowLine(50,0)(80,0){7}
\Vertex(50,0){3}
\Text(0,0)[]{$A_\mu{}^i_j(k)$}
\Text(105,0)[]{$\bar{\psi}^{bk}(q_1)$}
\end{picture}\hspace{30pt}
&=&
V(A_\mu{}^i_j(k),\bar{\psi}^{bk}(q_1))
\\&&\hspace{-10em}
=-g\int_{q_2}\tilde{\delta}(q_1+q_2+k)\,\gamma^\mu{}^a_b\,
  \delta^j_{k}\,\tilde{\psi}_{ai}(q_2)\,
  e^{\frac{i}{2}q_1\theta{}q_2}
\\
%---------------------------------------------------
\begin{picture}(120,20)(0,6)
\SetOffset(25,10)
\Photon(20,0)(50,0){3}{2.5}
\DashArrowLine(80,0)(50,0){7}
\Vertex(50,0){3}
\Text(0,0)[]{$A_\mu{}^i_j(k)$}
\Text(105,0)[]{$\psi_{al}(q_2)$}
\end{picture}\hspace{30pt}
&=&
V(A_\mu{}^i_j(k),\psi_{al}(q_2))
\\&&\hspace{-10em}
=g\int_{q_1}\tilde{\delta}(q_1+q_2+k)\,\gamma^\mu{}^a_b\,
  \delta^{l_2}_{i}\,\tilde{\bar{\psi}}{}^{bj}(q_1)\,
  e^{\frac{i}{2}q_1\theta{}q_2}
\\
%---------------------------------------------------
\begin{picture}(120,20)(0,6)
\SetOffset(25,10)
\Photon(20,0)(50,0){3}{2.5}
\SetWidth{1.4}
\DashArrowLine(50,0)(80,0){7}
\SetWidth{0.5}
\Vertex(50,0){3}
\Text(0,0)[]{$A_\mu{}^i_j(k)$}
\Text(105,0)[]{$\bar{\Psi}^b{}^{k_1}_{l_1}(Q_1)$}
\end{picture}\hspace{30pt}
&=&
V(A_\mu{}^i_j(k),\bar{\Psi}^b{}^{k_1}_{l_1}(Q_1))
\\&&\hspace{-10em}
=q\int_{Q_2}\tilde{\delta}(Q_1+Q_2+k)\,\gamma^\mu{}^a_b
    \left[\,
    \delta^{l_1}_{i}\tilde{\Psi}{}_a{}^{j}_{k_1}(Q_2)
    e^{-\frac{i}{2}Q_1\theta{}Q_2}
    -
    \delta^{j}_{k_1}
    \tilde{\Psi}{}_a{}^{l_1}_{i}(Q_2)
    e^{\frac{i}{2}Q_1\theta{}Q_2}
    \right]
\\
%---------------------------------------------------
\begin{picture}(120,20)(0,6)
\SetOffset(25,10)
\Photon(20,0)(50,0){3}{2.5}
\SetWidth{1.4}
\DashArrowLine(80,0)(50,0){7}
\SetWidth{0.5}
\Vertex(50,0){3}
\Text(0,0)[]{$A_\mu{}^i_j(k)$}
\Text(105,0)[]{$\Psi_a{}^{k_2}_{l_2}(Q_2)$}
\end{picture}\hspace{30pt}
&=&
V(A_\mu{}^i_j(k),\Psi_a{}^{k_2}_{l_2}(Q_2))
\\&&\hspace{-10em}
=q\int_{Q_1}\tilde{\delta}(Q_1+Q_2+k)\,\gamma^\mu{}^a_b
    \left[\,
    \delta^{l_2}_{i}\tilde{\bar{\Psi}}{}^b{}^{j}_{k_2}(Q_1)
    e^{\frac{i}{2}Q_1\theta{}Q_2}
    -
    \delta^{j}_{k_2}
    \tilde{\bar{\Psi}}{}^b{}^{l_2}_{i}(Q_1)
    e^{-\frac{i}{2}Q_1\theta{}Q_2}
    \right]
\\
%---------------------------------------------------
\begin{picture}(120,20)(0,6)
\SetOffset(25,10)
\Photon(20,0)(50,0){3}{2.5}
\ArrowLine(50,0)(80,0)
\Vertex(50,0){3}
\Text(0,0)[]{$A_\mu{}^i_j(k)$}
\Text(105,0)[]{$\phi^+{}^m(p_1)$}
\end{picture}\hspace{30pt}
&=&
V(A_\mu{}^i_j(k),\phi^+{}^m(p_1))
\\&&\hspace{-10em}=
g\int_{p_2}\tilde{\delta}(p_1+p_2+k)\,\delta^j_m\,\tilde{\phi}_i(p_2)
 \,(p_1^\mu-p_2^\mu)\,
 e^{\frac{i}{2}p_1\theta{}p_2}
\\&&\hspace{-10em}
+g^2\int_{p_2k_1}\tilde{\delta}(p_1+p_2+k+k_1)\,\delta^j_m\,
  \tilde{A}^\mu{}^n_i(k_1)\,
  \tilde{\phi}_n(p_2)
  e^{-\frac{i}{2}k\theta(k_1+p_2)-\frac{i}{2}k_1\theta{}p_2}
\\
%---------------------------------------------------
\begin{picture}(120,20)(0,6)
\SetOffset(25,10)
\Photon(20,0)(50,0){3}{2.5}
\ArrowLine(80,0)(50,0)
\Vertex(50,0){3}
\Text(0,0)[]{$A_\mu{}^i_j(k)$}
\Text(105,0)[]{$\phi_n(p_2)$}
\end{picture}\hspace{30pt}
&=&
V(A_\mu{}^i_j(k),\phi_n(p_2))
\\&&\hspace{-10em}=
g\int_{p_1}\tilde{\delta}(p_1+p_2+k)\,\delta^n_i\,\tilde{\phi}_*^j(p_1)
 \,(p_1^\mu-p_2^\mu)\,
 e^{\frac{i}{2}p_1\theta{}p_2}
\\&&\hspace{-10em}
+g^2\int_{p_1k_1}\tilde{\delta}(p_1+p_2+k+k_1)\,\delta^n_i\,
  \tilde{\phi}_*^m(p_1)
  \tilde{A}^\mu{}^j_m(k_1)\,
  e^{\frac{i}{2}k\theta(k_1+p_1)-\frac{i}{2}p_1\theta{}k_1}
\\
%---------------------------------------------------
\begin{picture}(120,20)(0,6)
\SetOffset(25,10)
\Photon(20,0)(50,0){3}{2.5}
\SetWidth{1.4}
\ArrowLine(50,0)(80,0)
\SetWidth{0.5}
\Vertex(50,0){3}
\Text(0,0)[]{$A_\mu{}^i_j(k)$}
\Text(108,0)[]{$\Phi^+{}^{m_1}_{n_1}(P_1)$}
\end{picture}\hspace{30pt}
&=&
V(A_\mu{}^i_j(k),\Phi^+{}^{m_1}_{n_1}(P_1))
\\&&\hspace{-10em}=
g\int_{P_2}\tilde{\delta}(P_1+P_2+k)\,(P_1^\mu-P_2^\mu)\,
  \left[\,
  \delta^{j}_{m_1}\tilde{\Phi}^{n_1}_{i}(P_2)
  e^{\frac{i}{2}P_1\theta{}P_2}
  -
  \delta^{n_1}_{i}\tilde{\Phi}^{j}_{m_1}(P_2)
  e^{-\frac{i}{2}P_1\theta{}P_2}
  \right]
\\&&\hspace{-10em}
+g^2\int_{P_2k_1}\tilde{\delta}(P_1+P_2+k_1+k)
  \left\{
  e^{\frac{i}{2}k\theta(k_1+P_2)}\delta^{n_1}_{i}\,
  \right.
\\&&\hspace{-9em}
    \times
    \left[
    \tilde{\Phi}^{k}_{m_1}(P_2)
    \tilde{A}^\mu{}^j_k(k_1)
    e^{-\frac{i}{2}P_2\theta{}k_1}
    -
    2
    \tilde{A}^\mu{}^k_{m_1}(k_1)
    \tilde{\Phi}^{j}_{k}(P_2)
    e^{\frac{i}{2}P_2\theta{}k_1}
    \right]
\\&&\hspace{-10em}
 + e^{-\frac{i}{2}k\theta(k_1+P_2)}\delta^{j}_{m_1}\,
\\&&\hspace{-9em}
    \left.
    \times
    \left[
    \tilde{A}^\mu{}^k_i(k_1)
    \tilde{\Phi}^{n_1}_{k}(P_2)
    e^{\frac{i}{2}P_2\theta{}k_1}
    -
    2
    \tilde{\Phi}^{k}_{i}(P_2)
    \tilde{A}^\mu{}^{n_1}_{k}(k_1)
    e^{-\frac{i}{2}P_2\theta{}k_1}
    \right]
  \right\}
\\
%---------------------------------------------------
\begin{picture}(120,20)(0,6)
\SetOffset(25,10)
\Photon(20,0)(50,0){3}{2.5}
\SetWidth{1.4}
\ArrowLine(80,0)(50,0)
\SetWidth{0.5}
\Vertex(50,0){3}
\Text(0,0)[]{$A_\mu{}^i_j(k)$}
\Text(105,0)[]{$\Phi^{m_2}_{n_2}(P_2)$}
\end{picture}\hspace{30pt}
&=&
V(A_\mu{}^i_j(k),\Phi^{m_2}_{n_2}(P_2))
\\&&\hspace{-10em}=
g\int_{P_1}\tilde{\delta}(P_1+P_2+k)\,(P_1^\mu-P_2^\mu)\,
  \left[\,
  \delta^{n_2}_{i}\tilde{\Phi}_*{}^{j}_{m_2}(P_1)
  e^{\frac{i}{2}P_1\theta{}P_2}
  -
  \delta^{j}_{m_2}\tilde{\Phi}_*{}^{n_2}_{i}(P_1)
  e^{-\frac{i}{2}P_1\theta{}P_2}
  \right]
\\&&\hspace{-10em}
+g^2\int_{P_1k_1}\tilde{\delta}(P_1+P_2+k_1+k)
  \left\{
  e^{-\frac{i}{2}k\theta(k_1+P_1)}\delta^{j}_{m_2}\,
  \right.
\\&&\hspace{-9em}
    \times
    \left[
    \tilde{A}^\mu{}^k_i(k_1)
    \tilde{\Phi}_*{}^{n_2}_{k}(P_1)
    e^{\frac{i}{2}P_1\theta{}k_1}
    -
    2
    \tilde{\Phi}_*{}^{k}_{i}(P_1)
    \tilde{A}^\mu{}^{n_2}_{k}(k_1)
    e^{-\frac{i}{2}P_1\theta{}k_1}
    \right]
\\&&\hspace{-10em}
 + e^{\frac{i}{2}k\theta(k_1+P_1)}\delta^{n_2}_{i}\,
\\&&\hspace{-9em}
    \left.
    \times
    \left[
    \tilde{\Phi}_*{}^{k}_{m_2}(P_1)
    \tilde{A}^\mu{}^j_k(k_1)
    e^{-\frac{i}{2}P_1\theta{}k_1}
    -
    2
    \tilde{A}^\mu{}^{k}_{m_2}(k_1)
    \tilde{\Phi}_*{}^{j}_{k}(P_1)
    e^{\frac{i}{2}P_1\theta{}k_1}
    \right]
  \right\}
%---------------------------------------------------
\\
\begin{picture}(120,20)(0,6)
\SetOffset(25,10)
\DashArrowLine(50,0)(20,0){2}
\DashArrowLine(80,0)(50,0){2}
\Vertex(50,0){3}
\Text(0,0)[]{$\bar{C}^{r_1}_{s_1}(l_1)$}
\Text(105,0)[]{$C^{r_2}_{s_2}(l_2)$}
\end{picture}\hspace{30pt}
&=&
V(\bar{C}^{r_1}_{s_1}(l_1),C^{r_2}_{s_2}(l_2))
=\frac{\delta_R}{\delta\tilde{C}{}^{r_2}_{s_2}(l_2)}
 \frac{\delta{}V}{\delta\tilde{\bar{C}}{}^{r_1}_{s_1}(l_1)}
\\&&\hspace{-10em}
 =g\int_{k}\tilde{\delta}(l_1+l_2+k)l_1^\mu
     \left[
     \delta^{s_2}_{r_1}\tilde{A}_\mu{}^{s_1}_{r_2}(k)
     e^{-\frac{i}{2}l_1\theta{}l_2}
     -
     \delta^{s_1}_{r_2}\tilde{A}_\mu{}^{s_2}_{r_1}(k)
     e^{\frac{i}{2}l_1\theta{}l_2}
     \right]
%---------------------------------------------------
\\
\begin{picture}(120,20)(0,6)
\SetOffset(25,10)
\DashArrowLine(20,0)(50,0){7}
\DashArrowLine(50,0)(80,0){7}
\Vertex(50,0){3}
\Text(0,0)[]{$\psi_{al}(q_2)$}
\Text(105,0)[]{$\bar{\psi}^{bk}(q_1)$}
\end{picture}\hspace{30pt}
&=&
V(\psi_{al}(q_2),\bar{\psi}^{bk}(q_1))
=
 \frac{\delta_R}{\delta\tilde{\bar\psi}{}^{bk}(q_1)}
 \frac{\delta_RV}{\tilde{\psi}{}_{al}(q_2)}
\\&&\hspace{-10em}=
 g\int_k\tilde{\delta}(q_1+q_2+k)\,\gamma^\mu{}^a_b
 \tilde{A}_\mu{}^l_k(k)
 e^{-\frac{i}{2}q_1\theta{}k}
%---------------------------------------------------
\\
\begin{picture}(120,20)(0,6)
\SetOffset(25,10)
\DashArrowLine(20,0)(50,0){7}
\SetWidth{1.4}
\DashArrowLine(50,0)(80,0){7}
\SetWidth{0.5}
\Vertex(50,0){3}
\Text(0,0)[]{$\psi_{al}(q_2)$}
\Text(105,0)[]{$\bar{\Psi}^b{}^{k_1}_{l_1}(Q_1)$}
\end{picture}\hspace{30pt}
&=&
V(\bar{\Psi}^b{}^{k_1}_{l_1}(Q_1),\psi_{al}(q_2))
\\&&\hspace{-10em}=
h^*\int_{p_1}\tilde{\delta}(p_1+Q_1+q_2)\,
  \delta^a_b\,
  \delta^{l}_{k_1}
  \tilde{\phi}_*{}^{l_1}(p_1)
  e^{-\frac{i}{2}q_2\theta{}p_1}
%---------------------------------------------------
\\
\begin{picture}(120,20)(0,6)
\SetOffset(25,10)
\DashArrowLine(20,0)(50,0){7}
\ArrowLine(50,0)(80,0)
\Vertex(50,0){3}
\Text(0,0)[]{$\psi_{al}(q_2)$}
\Text(105,0)[]{$\phi^+{}^m(p_1)$}
\end{picture}\hspace{30pt}
&=&
V(\psi_{al}(q_2),\phi^+{}^m(p_1))
\\&&\hspace{-10em}=
h^*\int_{Q_1}\tilde{\delta}(p_1+Q_1+q_2)\,
 \tilde{\bar\Psi}{}^{al}_{m}(Q_1)
 e^{\frac{i}{2}p_1\theta{}q_2}
%---------------------------------------------------
\\
\begin{picture}(120,20)(0,6)
\SetOffset(25,10)
\ArrowLine(20,0)(50,0)
\DashArrowLine(50,0)(80,0){7}
\Vertex(50,0){3}
\Text(0,0)[]{$\phi_{n}(p_2)$}
\Text(105,0)[]{$\bar{\psi}^{bk}(q_1)$}
\end{picture}\hspace{30pt}
&=&
V(\phi_{n}(p_2),\bar{\psi}^{bk}(q_1))
\\&&\hspace{-10em}=
h\int_{Q_2}\tilde{\delta}(q_1+Q_2+p_2)\,
 \tilde{\Psi}_b{}^n_k(Q_2)
 e^{\frac{i}{2}p_2\theta{}Q_2}
%---------------------------------------------------
\\
\begin{picture}(120,20)(0,6)
\SetOffset(25,10)
\SetWidth{1.4}
\DashArrowLine(20,0)(50,0){7}
\SetWidth{0.5}
\DashArrowLine(50,0)(80,0){7}
\Vertex(50,0){3}
\Text(-5,0)[]{$\Psi_{a}{}^{k_2}_{l_2}(Q_2)$}
\Text(105,0)[]{$\bar{\psi}^{bk}(q_1)$}
\end{picture}\hspace{30pt}
&=&
V(\bar{\psi}^{bk}(q_1),\Psi_{a}{}^{k_2}_{l_2}(Q_2))
\\&&\hspace{-10em}=
h\int_{p_2}\tilde{\delta}(q_1+Q_2+p_2)\,
  \delta^a_b\,
  \delta^{l_2}_{k}\,
  \tilde{\phi}_{k_2}(p_2)
  e^{\frac{i}{2}q_1\theta{}p_{2}}
%---------------------------------------------------
\\
\begin{picture}(120,20)(0,6)
\SetOffset(25,10)
\SetWidth{1.4}
\DashArrowLine(20,0)(50,0){7}
\DashArrowLine(50,0)(80,0){7}
\SetWidth{0.5}
\Vertex(50,0){3}
\Text(-5,0)[]{$\Psi_{a}{}^{k_2}_{l_2}(Q_2)$}
\Text(105,0)[]{$\bar{\Psi}^b{}^{k_1}_{l_1}(Q_1)$}
\end{picture}\hspace{30pt}
&=&
V(\Psi_{a}{}^{k_2}_{l_2}(Q_2),\bar{\Psi}^b{}^{k_1}_{l_1}(Q_1))
=
 \frac{\delta_R}{\delta\tilde{\bar\Psi}{}^b{}^{k_1}_{l_1}(Q_1)}
 \frac{\delta_RV}{\delta\tilde{\Psi}_a{}^{k_2}_{l_2}(Q_2)}
\\&&\hspace{-10em}=
 g\int_k\tilde{\delta}(Q_1+Q_2+k)
 \,\gamma^\mu{}^a_b
 \left[
 \delta^{l_1}_{k_2}
 \tilde{A}_\mu{}^{l_2}_{k_1}(k)
 e^{-\frac{i}{2}Q_1\theta{}k}
 -
 \delta^{l_2}_{k_1}
 \tilde{A}_\mu{}^{l_1}_{k_2}(k)
 e^{\frac{i}{2}Q_1\theta{}k}
 \right]
%---------------------------------------------------
\\
\begin{picture}(120,20)(0,6)
\SetOffset(25,10)
\SetWidth{1.4}
\DashArrowLine(20,0)(50,0){7}
\SetWidth{0.5}
\ArrowLine(80,0)(50,0)
\Vertex(50,0){3}
\Text(-5,0)[]{$\Psi_{a}{}^{k_2}_{l_2}(Q_2)$}
\Text(105,0)[]{$\phi_n(p_2)$}
\end{picture}\hspace{30pt}
&=&
V(\Psi_{a}{}^{k_2}_{l_2}(Q_2),\phi_n(p_2))
\\&&\hspace{-10em}=
h\int_{q_1}\tilde{\delta}(q_1+Q_2+p_2)\,
 \delta^{n}_{k_2}\,
 \tilde{\bar\psi}{}^{al_2}(q_1)
 e^{-\frac{i}{2}p_2\theta{}q_1}
%---------------------------------------------------
\\
\begin{picture}(120,20)(0,6)
\SetOffset(25,10)
\ArrowLine(50,0)(20,0)
\SetWidth{1.4}
\DashArrowLine(50,0)(80,0){7}
\SetWidth{0.5}
\Vertex(50,0){3}
\Text(-2,0)[]{$\phi^+{}^m(p_1)$}
\Text(105,0)[]{$\bar{\Psi}^b{}^{k_1}_{l_1}(Q_1)$}
\end{picture}\hspace{30pt}
&=&
V(\phi^+{}^m(p_1),\bar{\Psi}^b{}^{k_1}_{l_1}(Q_1))
\\&&\hspace{-10em}=
h^*\int_{q_2}\tilde{\delta}(p_1+Q_1+q_2)\,
 \delta^{l_1}_{m}\,
 \tilde{\psi}{}_{bk_1}(q_2)
 e^{\frac{i}{2}p_1\theta{}q_2}
%---------------------------------------------------
\\
\begin{picture}(120,20)(0,6)
\SetOffset(25,10)
\ArrowLine(50,0)(20,0)
\SetWidth{1.4}
\ArrowLine(80,0)(50,0)
\SetWidth{0.5}
\Vertex(50,0){3}
\Text(-2,0)[]{$\phi^+{}^m(p_1)$}
\Text(105,0)[]{$\Phi^{m_2}_{n_2}(P_2)$}
\end{picture}\hspace{30pt}
&=&
V(\phi^+{}^m(p_1),\Phi^{m_2}_{n_2}(P_2))
\\&&\hspace{-10em}=
-f_{a}\int_{p_2P_1}\tilde{\delta}(p_1+p_2+P_1+P_2)\,
 \delta^{n_2}_{m}\,
 \tilde{\Phi}_*{}^{k}_{m_2}(P_1)\,
 \tilde{\phi}_k(p_2)
 e^{\frac{i}{2}p_1\theta(p_2+P_1)-\frac{i}{2}P_1\theta{}p_2}
%---------------------------------------------------
\\
\begin{picture}(120,20)(0,6)
\SetOffset(25,10)
\ArrowLine(50,0)(20,0)
\SetWidth{1.4}
\ArrowLine(50,0)(80,0)
\SetWidth{0.5}
\Vertex(50,0){3}
\Text(-2,0)[]{$\phi^+{}^m(p_1)$}
\Text(105,0)[]{$\Phi^+{}^{m_1}_{n_1}(P_1)$}
\end{picture}\hspace{30pt}
&=&
V(\phi^+{}^m(p_1),\Phi^+{}^{m_1}_{n_1}(P_1))
\\&&\hspace{-10em}=
-f_{b}\int_{p_2P_2}
 \tilde{\delta}(p_1+p_2+P_1+P_2)\,
 \delta^{n_1}_{m}\,
 \tilde{\Phi}^{k}_{m_1}(P_2)
 \tilde{\phi}_{k}(p_2)
 e^{\frac{i}{2}p_1\theta(p_2+P_2)+\frac{i}{2}p_2\theta{}P_2}
%---------------------------------------------------
\\
\begin{picture}(120,20)(0,6)
\SetOffset(25,10)
\ArrowLine(20,0)(50,0)
\ArrowLine(50,0)(80,0)
\Vertex(50,0){3}
\Text(0,0)[]{$\phi_n(p_2)$}
\Text(105,0)[]{$\phi^+{}^m(p_1)$}
\end{picture}\hspace{30pt}
&=&
V(\phi_n(p_2),\phi^+{}^m(p_1))
\\&&\hspace{-10em}=
 g\int_{k}\tilde{\delta}(p_1+p_2+k)\,(p_1^\mu-p_2^\mu)\,
   \tilde{A}_\mu{}^n_m(k)
   e^{-\frac{i}{2}p_1\theta{}k}
\\&&\hspace{-9em}
+g^2\int_{k_1k_2}\tilde{\delta}(p_1+p_2+k_1+k_2)\,
 \tilde{A}_\mu{}^k_m(k_1)
 \tilde{A}^\mu{}^n_k(k_2)
 e^{-\frac{i}{2}p_1\theta(k_1+k_2)-\frac{i}{2}k_1\theta{}k_2}
\\&&\hspace{-9em}
-\frac{2\lambda_1}{4!}\int_{p_3p_4}\tilde{\delta}(p_1+p_2+p_3+p_4)\,
 \left[\,
 \delta^n_m
 \tilde{\phi}_*^k(p_3)
 \tilde{\phi}_k(p_4)
 e^{\frac{i}{2}p_1\theta(p_3+p_4)-\frac{i}{2}p_3\theta{}p_4}
 \right.
\\&&\hspace{-5em}
 +
 \left.
 \tilde{\phi}_*^n(p_3)
 \tilde{\phi}_m(p_4)
 e^{-\frac{i}{2}p_1\theta(p_3+p_4)+\frac{i}{2}p_3\theta{}p_4}
 \right]
\\&&\hspace{-9em}
-f_{a}\int_{P_1P_2}\tilde{\delta}(p_1+p_2+P_1+P_2)\,
 \tilde{\Phi}{}^k_m(P_2)
 \tilde{\Phi}_*{}^n_k(P_1)
 e^{-\frac{i}{2}p_1\theta(P_1+P_2)+\frac{i}{2}P_1\theta{}P_2}
\\&&\hspace{-9em}
-f_{b}\int_{P_1P_2}\tilde{\delta}(p_1+p_2+P_1+P_2)\,
 \tilde{\Phi}_*{}^k_m(P_1)
 \tilde{\Phi}{}^n_k(P_2)
 e^{-\frac{i}{2}p_1\theta(P_1+P_2)-\frac{i}{2}P_1\theta{}P_2}
%---------------------------------------------------
\\
\begin{picture}(120,20)(0,6)
\SetOffset(25,10)
\ArrowLine(20,0)(50,0)
\SetWidth{1.4}
\ArrowLine(80,0)(50,0)
\SetWidth{0.5}
\Vertex(50,0){3}
\Text(0,0)[]{$\phi_n(p_2)$}
\Text(105,0)[]{$\Phi^{m_2}_{n_2}(P_2)$}
\end{picture}\hspace{30pt}
&=&
V(\phi_n(p_2),\Phi^{m_2}_{n_2}(P_2))
\\&&\hspace{-10em}=
-f_{b}\int_{p_1P_1}\tilde{\delta}(p_1+p_2+P_1+P_2)\,
 \delta^{n}_{m_2}
 \tilde{\phi}_*{}^{k}(p_1)
 \tilde{\Phi}_*{}^{n_2}_{k}(P_1)
 e^{-\frac{i}{2}p_2\theta(p_1+P_1)-\frac{i}{2}p_1\theta{}P_1}
%---------------------------------------------------
\\
\begin{picture}(120,20)(0,6)
\SetOffset(25,10)
\ArrowLine(20,0)(50,0)
\SetWidth{1.4}
\ArrowLine(50,0)(80,0)
\SetWidth{0.5}
\Vertex(50,0){3}
\Text(0,0)[]{$\phi_n(p_2)$}
\Text(107,0)[]{$\Phi^+{}^{m_1}_{n_1}(P_1)$}
\end{picture}\hspace{30pt}
&=&
V(\phi_n(p_2),\Phi^+{}^{m_1}_{n_1}(P_1))
\\&&\hspace{-10em}=
-f_{a}\int_{p_1P_2}\tilde{\delta}(p_1+p_2+P_1+P_2)\,
 \delta^n_{m_1}
 \tilde{\phi}_*^k(p_1)
 \tilde{\Phi}^{n_1}_k(P_2)
 e^{-\frac{i}{2}p_2\theta(p_1+P_2)-\frac{i}{2}p_1\theta{}P_2}
%---------------------------------------------------
\\
\begin{picture}(120,20)(0,6)
\SetOffset(25,10)
\SetWidth{1.4}
\ArrowLine(20,0)(50,0)
\ArrowLine(80,0)(50,0)
\SetWidth{0.5}
\Vertex(50,0){3}
\Text(-2,0)[]{$\Phi^{m_2}_{n_2}(P_2)$}
\Text(105,0)[]{$\Phi^{m_4}_{n_4}(P_4)$}
\end{picture}\hspace{30pt}
&=&
V(\Phi^{m_2}_{n_2}(P_2),\Phi^{m_4}_{n_4}(P_4))
\\&&\hspace{-10em}=
-\frac{\lambda_{2b}}{4!}
    \int_{P_1P_3}
    \tilde{\delta}(P_1+P_2+P_3+P_4)\,
    \delta^{n_4}_{m_2}\,
 \tilde{\Phi}_*{}^{k}_{m_4}(P_1)
 \tilde{\Phi}_*{}^{n_2}_{k}(P_3)
 e^{\frac{i}{2}P_2\theta(P_1+P_3)-\frac{i}{2}P_1\theta{}P_3}
\\&&\hspace{-10em}\phantom{=}
-\frac{\lambda_{2b}}{4!}\int_{P_1P_3}\tilde{\delta}(P_1+P_2+P_3+P_4)\,
 \delta^{n_2}_{m_4}\,
 \tilde{\Phi}_*{}^{k}_{m_2}(P_1)
 \tilde{\Phi}_*{}^{n_4}_{k}(P_3)
 e^{-\frac{i}{2}P_2\theta(P_1+P_3)-\frac{i}{2}P_1\theta{}P_3}
\\&&\hspace{-10em}\phantom{=}
-\frac{\lambda_{3}}{3!}\int_{P_6P_8}\tilde{\delta}(P_2+P_4+P_6+P_8)\,
 \delta^{n_4}_{m_2}\,
 \tilde{\Phi}{}^{k}_{m_4}(P_6)
 \tilde{\Phi}{}^{n_2}_{k}(P_8)
 e^{\frac{i}{2}P_2\theta(P_6+P_8)-\frac{i}{2}P_6\theta{}P_8}
\\&&\hspace{-10em}\phantom{=}
-\frac{\lambda_{3}}{3!}\int_{P_6P_8}\tilde{\delta}(P_2+P_4+P_6+P_8)\,
 \delta^{n_2}_{m_4}\,
 \tilde{\Phi}{}^{k}_{m_2}(P_6)
 \tilde{\Phi}{}^{n_4}_{k}(P_8)
 e^{-\frac{i}{2}P_2\theta(P_6+P_8)-\frac{i}{2}P_6\theta{}P_8}
%---------------------------------------------------
\\
\begin{picture}(120,20)(0,6)
\SetOffset(25,10)
\SetWidth{1.4}
\ArrowLine(20,0)(50,0)
\ArrowLine(50,0)(80,0)
\SetWidth{0.5}
\Vertex(50,0){3}
\Text(-2,0)[]{$\Phi^{m_2}_{n_2}(P_2)$}
\Text(105,0)[]{$\Phi^+{}^{m_1}_{n_1}(P_1)$}
\end{picture}\hspace{30pt}
&=&
V(\Phi^{m_2}_{n_2}(P_2),\Phi^+{}^{m_1}_{n_1}(P_1))
\\&&\hspace{-10em}=
g\int_{k}\tilde{\delta}(P_1+P_2+k)\,(P_1^\mu-P_2^\mu)
 \left[\,
 \delta^{n_1}_{m_2}\,
 \tilde{A}_\mu{}^{n_2}_{m_1}(k)
 e^{-\frac{i}{2}P_1\theta{}k}
 -
 \delta^{n_2}_{m_1}\,
 \tilde{A}_\mu{}^{n_1}_{m_2}(k)
 e^{\frac{i}{2}P_1\theta{}k}
 \right]
\\&&\hspace{-10em}\phantom{=}
+g^2\int_{k_1k_2}\tilde{\delta}(P_1+P_2+k_1+k_2)\,
  e^{-\frac{i}{2}k_1\theta{}k_2}\,
  \left[\,
  \delta^{n_2}_{m_1}\,
  \tilde{A}_\mu{}^{k}_{m_2}(k_1)
  \tilde{A}^\mu{}^{n_1}_{k}(k_2)
  e^{\frac{i}{2}P_1\theta{}k}
  \right.
\\&&\hspace{-5em}\phantom{=}
  +
  \left.
  \delta^{n_1}_{m_2}\,
  \tilde{A}_\mu{}^{k}_{m_1}(k_1)
  \tilde{A}^\mu{}^{n_2}_{k}(k_2)
  e^{-\frac{i}{2}P_1\theta{}k}
  \right]
\\&&\hspace{-10em}\phantom{=}
-\frac{2\lambda_{2a}}{4!}
  \int_{P_3P_4}\tilde{\delta}(P_1+P_2+P_3+P_4)\,
  \left[\,
  \delta^{n_1}_{m_2}
  \tilde{\Phi}{}^{k}_{m_1}(P_4)
  \tilde{\Phi}_*{}^{n_2}_{k}(P_3)
  e^{-\frac{i}{2}P_1\theta(P_3+P_4)+\frac{i}{2}P_3\theta{}P_4}
  \right.
\\&&\hspace{-5em}\phantom{=}
  +
  \left.
  \delta^{n_2}_{m_1}
  \tilde{\Phi}_*{}^{k}_{m_2}(P_3)
  \tilde{\Phi}{}^{n_1}_{k}(P_4)
  e^{\frac{i}{2}P_1\theta(P_3+P_4)-\frac{i}{2}P_3\theta{}P_4}
  \right]
\\&&\hspace{-10em}\phantom{=}
-\frac{\lambda_{2b}}{4!}
  \int_{P_3P_4}\tilde{\delta}(P_1+P_2+P_3+P_4)\,
  \left[\,
  \delta^{n_2}_{m_1}
  \tilde{\Phi}_*{}^{n_1}_{k}(P_3)
  \tilde{\Phi}{}^{k}_{m_2}(P_4)
  e^{\frac{i}{2}P_1\theta(P_3+P_4)+\frac{i}{2}P_3\theta{}P_4}
  \right.
\\&&\hspace{-5em}\phantom{=}
  +
  \left.
  \delta^{n_1}_{m_2}
  \tilde{\Phi}_*{}^{k}_{m_1}(P_3)
  \tilde{\Phi}{}^{n_2}_{k}(P_4)
  e^{-\frac{i}{2}P_1\theta(P_3+P_4)-\frac{i}{2}P_3\theta{}P_4}
  \right]
\\&&\hspace{-10em}\phantom{=}
-f_a\int_{p_1p_2}\tilde{\delta}(p_1+p_2+P_1+P_2)\,
  \delta^{n_1}_{m_2}
  \tilde{\phi}_*{}^{n_2}(p_1)
  \tilde{\phi}_{m_1}(p_2)
  e^{-\frac{i}{2}P_1\theta(p_1+p_2)+\frac{i}{2}p_1\theta{}p_2}
\\&&\hspace{-10em}\phantom{=}
-f_b\int_{p_1p_2}\tilde{\delta}(p_1+p_2+P_1+P_2)\,
  \delta^{n_2}_{m_1}
  \tilde{\phi}_*{}^{n_1}(p_1)
  \tilde{\phi}_{m_2}(p_2)
  e^{\frac{i}{2}P_1\theta(p_1+p_2)+\frac{i}{2}p_1\theta{}p_2}
%---------------------------------------------------
\\
\begin{picture}(120,20)(0,6)
\SetOffset(25,10)
\SetWidth{1.4}
\ArrowLine(50,0)(20,0)
\ArrowLine(50,0)(80,0)
\SetWidth{0.5}
\Vertex(50,0){3}
\Text(-5,0)[]{$\Phi^+{}^{m_1}_{n_1}(P_1)$}
\Text(105,0)[]{$\Phi^+{}^{m_3}_{n_3}(P_3)$}
\end{picture}\hspace{30pt}
&=&
V(\Phi^+{}^{m_1}_{n_1}(P_1),\Phi^+{}^{m_3}_{n_3}(P_3))
\\&&\hspace{-10em}=
-\frac{\lambda_{2b}}{4!}\int_{P_2P_4}\tilde{\delta}(P_1+P_2+P_3+P_4)\,
 \delta^{n_3}_{m_1}\,
 \tilde{\Phi}{}^{k}_{m_3}(P_2)\,
 \tilde{\Phi}{}^{n_1}_{k}(P_4)\,
 e^{\frac{i}{2}P_1\theta(P_2+P_4)-\frac{i}{2}P_2\theta{}P_4}
\\&&\hspace{-10em}\phantom{=}
-\frac{\lambda_{2b}}{4!}\int_{P_2P_4}\tilde{\delta}(P_1+P_2+P_3+P_4)\,
 \delta^{n_1}_{m_3}\,
 \tilde{\Phi}{}^{k}_{m_1}(P_2)\,
 \tilde{\Phi}{}^{n_3}_{k}(P_4)\,
 e^{-\frac{i}{2}P_1\theta(P_2+P_4)-\frac{i}{2}P_2\theta{}P_4}
\\&&\hspace{-10em}\phantom{=}
-\frac{\lambda^*_3}{3!}
      \int_{P_5P_7}
     \tilde{\delta}(P_1+P_3+P_5+P_7)\,
 \delta^{n_3}_{m_1}
 \tilde{\Phi}_*{}^{k}_{m_3}(P_5)\,
 \tilde{\Phi}_*{}^{n_1}_{k}(P_7)\,
 e^{\frac{i}{2}P_1\theta(P_5+P_7)-\frac{i}{2}P_5\theta{}P_7}
\\&&\hspace{-10em}\phantom{=}
-\frac{\lambda^*_3}{3!}
      \int_{P_5P_7}
      \tilde{\delta}(P_1+P_3+P_5+P_7)\,
 \delta^{n_1}_{m_3}
 \tilde{\Phi}_*{}^{k}_{m_1}(P_5)\,
 \tilde{\Phi}_*{}^{n_3}_{k}(P_7)\,
 e^{-\frac{i}{2}P_1\theta(P_5+P_7)-\frac{i}{2}P_5\theta{}P_7}
%---------------------------------------------------
\end{eqnarray*}

%\newpage
%%%%%%%%%%%%%%%%%%%%%%%%%%%%%%%%%%%%%%%%%%%%%%%%%%%%%%%%%%%
%%%%%%%%%%%%%%%%%%%%%%%%%%%%%%%%%%%%%%%%%%%%%%%%%%%%%%%%%%%
%%%%%%%%%%%%%%%%%%%%%%%%%%%%%%%%%%%%%%%%%%%%%%%%%%%%%%%%%%%
%%%%%%%%%%%%%%%%%%%%%%%%%%%%%%%%%%%%%%%%%%%%%%%%%%%%%%%%%%%

%%%%%%%%%%%%%%%%%%%%%%%%%%%%%%%%%%%%%%%%%%%%%%%%%%%%%%%%%%%%%%%
\end{document}